\documentclass[aps,prd,nofootinbib,floatfix,superscriptaddress,secnumarabic,preprintnumbers,longbibliography,10pt]{revtex4-2}
\usepackage{amsmath}
\usepackage{graphicx}
\usepackage[dvipsnames]{xcolor}
\usepackage{amsfonts}
\usepackage{hyperref}
\usepackage{slashed}
\usepackage{ytableau}
\usepackage{relsize}
\usepackage[normalem]{ulem}
\usepackage{mathtools}

\DeclareFontFamily{U}{mathx}{}
\DeclareFontShape{U}{mathx}{m}{n}{<-> mathx10}{}
\DeclareSymbolFont{mathx}{U}{mathx}{m}{n}
\DeclareMathAccent{\widehat}{0}{mathx}{"70}
\DeclareMathAccent{\widecheck}{0}{mathx}{"71}

\DeclareRobustCommand{\svdots}{
  \vbox{%
    \baselineskip=0.233333\normalbaselineskip
    \lineskiplimit=0pt
    \hbox{.}\hbox{.}\hbox{.}%
    \kern-0.2\baselineskip
  }%
}
\newcommand{\MSbar}{{\overline{\rm MS}}}
\newcommand{\be}{\begin{equation}}
\newcommand{\ee}{\end{equation}}
\newcommand{\bea}{\begin{eqnarray}}
\newcommand{\eea}{\end{eqnarray}}

\begin{document}

\title{Classification of four-quark operators with $\Delta F\le 2$ under flavor symmetry and their renormalization in a gauge-invariant scheme}
\author{G.~Spanoudes}
\email[]{spanoudes.gregoris@ucy.ac.cy}
\affiliation{Department of Physics, University of Cyprus, Nicosia, CY-1678, Cyprus}

\author{M.~Costa}
\email[]{marios.costa@cut.ac.cy}
\affiliation{Current address: Department of Mechanical Engineering and Material Science and Engineering, Cyprus University of Technology, Limassol, CY-3036, Cyprus}
\affiliation{Rinnoco Ltd, Limassol, CY-3047, Cyprus}

\author{K.~Mitsidi}
\email[]{mitsidi.kyproula@ucy.ac.cy}
\affiliation{Department of Physics, University of Cyprus, Nicosia, CY-1678, Cyprus}

\author{H.~Panagopoulos}
\email[]{haris@ucy.ac.cy}
\affiliation{Department of Physics, University of Cyprus, Nicosia, CY-1678, Cyprus}

\begin{abstract}
In this paper we study a complete set of scalar and pseudoscalar four-quark operators, with a particular emphasis on their renormalization within a Gauge-Invariant Renormalization Scheme (GIRS). We focus on operators that do not mix with lower-dimensional operators by virtue of their transformation properties under the flavor-symmetry group. This class includes all $\Delta F = 2$ operators, as well as their partners that transform under the same irreducible representations of the flavor group. These encompass a substantial subset of $\Delta F = 1$ and $\Delta F = 0$ operators. The present analysis provides a detailed classification of all four-quark operators, exploring their Fierz identities, symmetry properties, and mixing patterns. Different variants of GIRS are explored, including a democratic version that treats all mixing operators uniformly. For selected variants, which exhibit smaller mixing effects, we present the conversion matrices from GIRS to the $\overline{\text{MS}}$ scheme at next-to-leading order. 
\end{abstract}

\maketitle

\section{Introduction}
\label{sec:Intro}

The study of four-quark operators plays a central role in understanding flavor-changing processes, CP violation, and in probing possible extensions of the Standard Model (SM). These operators naturally arise in effective field theories that describe low-energy hadronic interactions, such as the weak effective Hamiltonian, and in scenarios involving lepton flavor violation and other forms of Beyond-the-Standard-Model (BSM) physics~\cite{FlavourLatticeAveragingGroupFLAG:2024oxs}. A crucial aspect of these studies is the proper renormalization of composite operators, as their matrix elements depend on the renormalization scale and scheme. Understanding their renormalization properties, especially in nonperturbative regimes, is crucial for matching lattice QCD calculations to continuum schemes, and for providing accurate predictions of physical observables. Over the past two decades, considerable effort has been devoted to the perturbative and nonperturbative renormalization of such operators using different discretizations on the lattice (see, e.g.,~\cite{Ciuchini:1997bw, Buchalla:1995vs, Donini:1999nn, Buras:2000if, Aoki:2003uf,Frezzotti:2004wz, Guagnelli:2005zc, Aoki:2006br, Nakamura:2006zx, Constantinou:2010zs, Boyle:2012qb, Taniguchi:2012xm, ETM:2013opv, Carrasco:2015pra, Dimopoulos:2018zef} and references therein), with notable advancements in more recent works ~\cite{Martinelli:2022xir, Martinelli:2022tte, Constantinou:2024wdb, Lin:2024mws, DiCarlo:2025mnm, Frezzotti:2025hif}. 

The matrix elements of four-quark operators within lattice QCD are essential for connecting experimental observables to fundamental parameters of the SM, such as the elements of the CKM matrix. Among the most prominent examples is the so-called bag parameter $B_K$, which parametrizes the hadronic matrix element relevant for $K^0 - \overline{K}^0$ mixing. This parameter is a key input for determining indirect CP violation in the kaon sector and contributes significantly to global fits of the unitarity triangle~\cite{Gupta:1992bd, Becirevic:2001re, Aoki:2010pe, ETM:2010ubf, Aebischer:2020dsw, Suzuki:2020zue}. Beyond kaons, four-quark operators are equally relevant for $B$- and $D$-meson mixing, where their matrix elements enter in the prediction of mass differences and mixing-induced CP asymmetries~\cite{Lenz:2014jha, Neubert:1996we}. In addition, such operators appear in many scenarios of physics beyond the SM, where they can induce flavor-changing effects not allowed within the SM or modify the strength of existing transitions~\cite{Gabbiani:1996hi}. As experimental precision continues to improve, particularly with new results from LHCb~\cite{LHCb:2025ray} and Belle II, matching this progress on the theoretical side requires a deeper understanding of four-quark operators from first principles. Lattice QCD provides a systematic framework to compute the relevant matrix elements nonperturbatively, with increasing control over continuum extrapolations, renormalization effects, and systematic uncertainties~\cite{Ishikawa:2011dd, DiPierro:1998ty, DiPierro:1999tb, Gimenez:1998mw, Ishizuka:2018qbn}.

To extract physical predictions, the matrix elements of four-quark operators computed in Lattice QCD must be consistently matched to Wilson coefficients evaluated in a given continuum renormalization scheme, most commonly the $\overline{\text{MS}}$ scheme~\cite{Martinelli:1994ty}. However, since the $\overline{\text{MS}}$ scheme is defined in the continuum, it is not directly accessible in lattice computations. To overcome this limitation, intermediate regularization-independent renormalization schemes are typically employed on the lattice, followed by the application of conversion factors to connect to the $\overline{\text{MS}}$ scheme. These conversion factors, being independent of the regulators, can be computed in dimensional regularization. One such intermediate scheme is the Gauge-Invariant Renormalization Scheme (GIRS)~\cite{Costa:2021pfu, Costa:2021iyv, Bergner:2022see, Constantinou:2024wdb}, which generalizes the coordinate-space ($X$-space) approach~\cite{Gimenez:2004me, Chetyrkin:2010dx, Cichy:2012is, Tomii:2018zix}. GIRS enables the nonperturbative renormalization of composite operators through the use of gauge-invariant correlators defined in position space. These methods are essential for bridging the gap between lattice regularizations and continuum perturbation theory, thereby enabling reliable phenomenological predictions and improved control over systematic uncertainties in the renormalization of four-quark operator matrix elements.

In this paper, we investigate scalar and pseudoscalar four-quark operators that can be renormalized within GIRS, developing further the framework recently applied to operators that change flavor numbers by two units ($\Delta F = 2$)~\cite{Constantinou:2024wdb}. Our focus is on operators whose transformation properties under the flavor symmetry group forbid mixing with lower-dimensional operators, thereby allowing for a cleaner and more controlled analysis of their renormalization and mixing patterns. We provide the theoretical foundation for a GIRS-based renormalization of these operators on the lattice, including a classification according to flavor symmetry representations, an exploration of Fierz identities, and a study of their transformation properties under parity and charge conjugation. In particular, we revisit and extend the classification of four-quark operators for a general number of flavors $N_f$, covering not only the $\Delta F = 2$ case, but also a broad class of $\Delta F = 1$ and $\Delta F = 0$ operators that reside in the same flavor representations.

The present paper is organized as follows. In Section~\ref{sec:Op_basis}, we define generic four-quark operators and classify them according to their transformation properties under the flavor symmetry group, making use of Fierz identities to reduce redundancies. Section~\ref{sec:(pseudo)scalar} focuses specifically on the scalar and pseudoscalar four-quark operators of interest. We discuss the mixing patterns of these operators and how flavor symmetry prevents mixing with lower-dimensional operators for specific representations of the flavor group. Additionally, by employing parity, charge conjugation and chiral transformations, we further decompose the mixing sets. Section~\ref{renorm} provides a brief review of the renormalization setup; it describes its application to the operators considered in this work using the GIRS scheme. In section~\ref{Results}, we construct renormalization conditions in GIRS and provide perturbative results for the conversion matrices to the $\MSbar$ scheme. Finally, in Section~\ref{sec:conclusions}, we summarize our findings and outline possible directions for future research.

\section{Operator setup}
\label{sec:Op_basis}

In this section, we define the complete set of four-quark operators relevant to our analysis, extending the operator set examined in our previous work~\cite{Constantinou:2024wdb}. We consider all representations of the flavor group that forbid mixing with lower-dimensional operators. This choice ensures a clean operator basis that is suitable for renormalization and matching calculations, free from contamination by lower-dimensional terms.

A four-quark operator has the general form:
\begin{equation}
({\cal O}_{\Gamma \Gamma'})^{f_1, f_2}_{f_3,f_4} \equiv \Big(\sum_{a} \sum_{\alpha,\beta} {\bar \psi}_{\alpha}^{a, f_1}(x)  \,\Gamma_{\alpha \beta} \,  \psi_{\beta, f_3}^{a}(x) \Big) \Big(\sum_{a'} \sum_{\alpha',\beta'} {\bar \psi}_{\alpha'}^{a',f_2}(x) \,\Gamma'_{\alpha' \beta'} \, \psi_{\beta',f_4}^{a'}(x) \Big),
\label{fourQ}
\end{equation}
where $\Gamma, \Gamma' \in \{\openone,\, \gamma_5,\, \gamma_\mu,\, \gamma_\mu \gamma_5,\,\sigma_{\mu\nu}, \,\sigma_{\mu\nu}\gamma_5\} \equiv \{S,P,V,A,T,\tilde T \}$ and $\sigma_{\mu\nu}=\frac{1}{2}[\gamma_\mu,\gamma_\nu]$; spinor indices are denoted by Greek letters ($\alpha$, $\beta$, $\alpha'$, $\beta'$), while color and flavor indices denoted by Latin letters ($a$, $a'$) and $f_i$, respectively. Throughout the paper, we adopt the convention that quantities with lower (upper) flavor indices transform according to the (anti-)fundamental representation of the flavor group. In order to classify the four-quark operators into irreducible representations of the $SU(N_f)$ flavor group, it is convenient to  denote operators with exchanged flavors of their quark fields ($f_3, f_4$) as~\cite{Donini:1999nn}: 
\begin{equation}
({\cal O}_{\Gamma \Gamma'}^F)^{f_1, f_2}_{f_3,f_4} \equiv ({\cal O}_{\Gamma \Gamma'})^{f_1, f_2}_{f_4,f_3} = \Big(\sum_{a} \sum_{\alpha,\beta} {\bar \psi}_{\alpha}^{a, f_1}(x)  \,\Gamma_{\alpha \beta} \,  {\psi_{\beta, f_4}^{a}}(x) \Big) \Big(\sum_{a'} \sum_{\alpha',\beta'} {\bar \psi}_{\alpha'}^{a',f_2}(x) \,\Gamma'_{\alpha' \beta'} \, {\psi_{\beta',f_3}^{a'}}(x) \Big).
\label{fourQF}
\end{equation}
The following relations between operators with exchanged flavor indices hold:
\begin{eqnarray}
({\cal O}_{\Gamma \Gamma'})^{f_2, f_1}_{f_3,f_4} &=& ({\cal O}_{\Gamma' \Gamma}^F)^{f_1, f_2}_{f_3,f_4}\ , \label{Exchange_rel1}\\
({\cal O}_{\Gamma \Gamma'})^{f_2, f_1}_{f_4,f_3} &=& ({\cal O}_{\Gamma' \Gamma})^{f_1, f_2}_{f_3,f_4}\ . \label{Exchange_rel2}
\end{eqnarray}
Making use of Fierz rearrangements, operators in Eqs.~\eqref{fourQ} and ~\eqref{fourQF} can be also expressed as linear combinations of those given in Eqs.~\eqref{fourQFD} and ~\eqref{fourQFDF}, respectively:
\begin{eqnarray}
({\cal O}_{\Gamma \Gamma'}^{F_D})^{f_1, f_2}_{f_3,f_4} &\equiv& \sum_{a, a'} \Big(\sum_{\alpha,\beta} {\bar \psi}_{\alpha}^{a, f_1}(x)  \,\Gamma_{\alpha \beta} \,  {\psi_{\beta, f_4}^{a'}}(x) \Big) \Big(\sum_{\alpha',\beta'} {\bar \psi}_{\alpha'}^{a',f_2}(x) \,\Gamma'_{\alpha' \beta'} \, {\psi_{\beta', f_3}^{a}}(x) \Big), \label{fourQFD} \\
({\cal O}_{\Gamma \Gamma'}^{F_D, F})^{f_1, f_2}_{f_3,f_4} &\equiv& \sum_{a, a'} \Big(\sum_{\alpha,\beta} {\bar \psi}_{\alpha}^{a, f_1}(x)  \,\Gamma_{\alpha \beta} \,  {\psi_{\beta, f_3}^{a'}}(x) \Big) \Big(\sum_{\alpha',\beta'} {\bar \psi}_{\alpha'}^{a',f_2}(x) \,\Gamma'_{\alpha' \beta'} \, {\psi_{\beta',f_4}^{a}}(x) \Big). 
\label{fourQFDF}
\end{eqnarray}
We will make this relation explicit in Section \ref{sec:(pseudo)scalar} for particular cases of interest. 

\subsection{Operator classification into irreducible representations of $SU(N_f)$ flavor group}
\label{OperatorClassification}
We focus on a theory with $N_f$ mass-degenerate quarks, where $SU(N_f)$ flavor symmetry holds. The symmetry is also preserved for theories with nondegenerate quarks when the chiral limit is taken. The latter case is typically considered when operator renormalization is studied. In these cases, the four-quark operators can be classified into specific sets that support different irreducible representations of the $SU(N_f)$ flavor group. We perform below the operator classification using group theory and Young tableaux. A similar investigation can be found in Ref. \cite{Donini:1999sf}. 

We first define operators with symmetrized or antisymmetrized quark and antiquark flavor indices, as follows: 
\begin{eqnarray}
   ({\cal O}^{+}_{\{\Gamma, \Gamma' \}})^{f_1, f_2}_{f_3,f_4} &\equiv& \frac{1}{2} \Big[({\cal O}_{\Gamma \Gamma'})^{f_1, f_2}_{f_3,f_4} + ({\cal O}_{\Gamma \Gamma'})^{f_2, f_1}_{f_3,f_4} + ({\cal O}_{\Gamma \Gamma'})^{f_1, f_2}_{f_4,f_3} + ({\cal O}_{\Gamma \Gamma'})^{f_2, f_1}_{f_4,f_3} \Big] = \frac{1}{2} \Big[({\cal O}_{\{\Gamma, \Gamma'\}})^{f_1, f_2}_{f_3,f_4} + ({\cal O}_{\{\Gamma, \Gamma'\}}^F)^{f_1, f_2}_{f_3,f_4}\Big], \label{Osym1} \\
   ({\cal O}^{-}_{\{\Gamma, \Gamma' \}})^{f_1, f_2}_{f_3,f_4} &\equiv& \frac{1}{2} \Big[({\cal O}_{\Gamma \Gamma'})^{f_1, f_2}_{f_3,f_4} - ({\cal O}_{\Gamma \Gamma'})^{f_2, f_1}_{f_3,f_4} - ({\cal O}_{\Gamma \Gamma'})^{f_1, f_2}_{f_4,f_3} + ({\cal O}_{\Gamma \Gamma'})^{f_2, f_1}_{f_4,f_3} \Big] = \frac{1}{2} \Big[({\cal O}_{\{\Gamma, \Gamma'\} })^{f_1, f_2}_{f_3,f_4} - ({\cal O}_{\{\Gamma, \Gamma'\}}^F)^{f_1, f_2}_{f_3,f_4}\Big], \\
   ({\cal O}^{+}_{[\Gamma, \Gamma' ]})^{f_1, f_2}_{f_3,f_4} &\equiv& \frac{1}{2} \Big[({\cal O}_{\Gamma \Gamma'})^{f_1, f_2}_{f_3,f_4} - ({\cal O}_{\Gamma \Gamma'})^{f_2, f_1}_{f_3,f_4} + ({\cal O}_{\Gamma \Gamma'})^{f_1, f_2}_{f_4,f_3} - ({\cal O}_{\Gamma \Gamma'})^{f_2, f_1}_{f_4,f_3} \Big] = \frac{1}{2} \Big[({\cal O}_{[\Gamma, \Gamma']})^{f_1, f_2}_{f_3,f_4} + ({\cal O}_{[\Gamma, \Gamma']}^F)^{f_1, f_2}_{f_3,f_4}\Big], \\
({\cal O}^{-}_{[\Gamma, \Gamma' ]})^{f_1, f_2}_{f_3,f_4} &\equiv& \frac{1}{2} \Big[({\cal O}_{\Gamma \Gamma'})^{f_1, f_2}_{f_3,f_4} + ({\cal O}_{\Gamma \Gamma'})^{f_2, f_1}_{f_3,f_4} - ({\cal O}_{\Gamma \Gamma'})^{f_1, f_2}_{f_4,f_3} - ({\cal O}_{\Gamma \Gamma'})^{f_2, f_1}_{f_4,f_3} \Big] = \frac{1}{2} \Big[({\cal O}_{[\Gamma, \Gamma']})^{f_1, f_2}_{f_3,f_4} - ({\cal O}_{[\Gamma, \Gamma']}^F)^{f_1, f_2}_{f_3,f_4}\Big], \label{Osym4}
\end{eqnarray}
where 
\begin{eqnarray}
(\mathcal{O}_{ \{\Gamma, \Gamma'\}})^{f_1, f_2}_{f_3, f_4} &\equiv& (\mathcal{O}_{\Gamma\Gamma' + \Gamma'\Gamma})^{f_1, f_2}_{f_3, f_4} \equiv (\mathcal{O}_{\Gamma \Gamma'})^{f_1, f_2}_{f_3, f_4} + (\mathcal{O}_{\Gamma' \Gamma})^{f_1, f_2}_{f_3, f_4}, \\
(\mathcal{O}_{[\Gamma, \Gamma']})^{f_1, f_2}_{f_3, f_4} &\equiv& (\mathcal{O}_{\Gamma\Gamma' - \Gamma'\Gamma})^{f_1, f_2}_{f_3, f_4} \equiv (\mathcal{O}_{\Gamma \Gamma'})^{f_1, f_2}_{f_3, f_4} -(\mathcal{O}_{\Gamma' \Gamma})^{f_1, f_2}_{f_3, f_4},
\end{eqnarray}
and Eqs. (\ref{Exchange_rel1}--\ref{Exchange_rel2}) have been used to obtain the r.h.s. of Eqs. (\ref{Osym1}--\ref{Osym4}). 

The original operators can be decomposed into 10 sets which support 10 irreducible representations of $SU(N_f)$:
    \begin{eqnarray}
    (\widecheck{\mathcal{O}}^{\pm}_{\{\Gamma, \Gamma'\}})^{f_1, f_2}_{f_3, f_4} &\equiv& ({\cal O}^{\pm}_{\{\Gamma, \Gamma'\}})^{f_1, f_2}_{f_3,f_4} - \frac{1}{(N_f \pm 2)} (\overline{\mathcal{O}}^{\pm}_{\{\Gamma, \Gamma'\}})^{f_1, f_2}_{f_3, f_4} - \frac{1}{N_f (N_f \pm 1)} (\widehat{\mathcal{O}}^{\pm}_{\{\Gamma, \Gamma'\}})^{f_1, f_2}_{f_3, f_4}, \label{Op1} \\
    (\overline{\mathcal{O}}^{\pm}_{\{\Gamma, \Gamma'\}})^{f_1, f_2}_{f_3, f_4} &\equiv& \sum_{f'} \Big( {{\delta}^{f_1}}_{f_3} ({\cal O}^{\pm}_{\{\Gamma, \Gamma'\}})^{f', f_2}_{f',f_4} + {\delta^{f_2}}_{f_3} ({\cal O}^{\pm}_{\{\Gamma, \Gamma'\}})^{f_1, f'}_{f',f_4} + {\delta^{f_1}}_{f_4} ({\cal O}^{\pm}_{\{\Gamma, \Gamma'\}})^{f', f_2}_{f_3,f'} + {\delta^{f_2}}_{f_4} ({\cal O}^{\pm}_{\{\Gamma, \Gamma'\}})^{f_1, f'}_{f_3,f'} \Big)  \nonumber \\
    && - \frac{2}{N_f} (\widehat{\mathcal{O}}^{\pm}_{\{\Gamma, \Gamma'\}})^{f_1, f_2}_{f_3, f_4}, \\
    (\widehat{\mathcal{O}}^{\pm}_{\{\Gamma, \Gamma'\}})^{f_1, f_2}_{f_3, f_4} &\equiv& ({\delta^{f_1}}_{f_3} {\delta^{f_2}}_{f_4} \pm {\delta^{f_1}}_{f_4} {\delta^{f_2}}_{f_3}) \sum_{f',f''} ({\cal O}^{\pm}_{\{\Gamma, \Gamma'\}})^{f', f''}_{f',f''}, \\    
    (\widecheck{\mathcal{O}}^{\pm}_{[\Gamma, \Gamma']})^{f_1, f_2}_{f_3, f_4} &\equiv& (\mathcal{O}^{\pm}_{[\Gamma, \Gamma']})^{f_1, f_2}_{f_3, f_4} - \frac{1}{N_f} (\overline{\mathcal{O}}^{\pm}_{[\Gamma, \Gamma']})^{f_1, f_2}_{f_3, f_4}, \label{Op4} \\
    (\overline{\mathcal{O}}^{\pm}_{[\Gamma, \Gamma']})^{f_1, f_2}_{f_3, f_4} &\equiv& \sum_{f'} \Big( {\delta^{f_1}}_{f_3} ({\cal O}^{\pm}_{[\Gamma, \Gamma']})^{f', f_2}_{f',f_4} + {\delta^{f_2}}_{f_3} ({\cal O}^{\pm}_{[\Gamma, \Gamma']})^{f_1, f'}_{f',f_4} + {\delta^{f_1}}_{f_4} ({\cal O}^{\pm}_{[\Gamma, \Gamma']})^{f', f_2}_{f_3,f'} + {\delta^{f_2}}_{f_4} ({\cal O}^{\pm}_{[\Gamma, \Gamma']})^{f_1, f'}_{f_3,f'} \Big), \label{Op5}
    \end{eqnarray}
$\widecheck{\mathcal{O}}^{\pm}$ are completely traceless, $\widehat{\mathcal{O}}^{\pm}$ are pure trace (thus flavor singlets), and $\overline{\mathcal{O}}^{\pm}$ support the adjoint representation of the flavor group. The corresponding Young tableaux for each representation are given in Fig.~\ref{Young_tableau}. Also, in Table~\ref{tab:dimensionalities}, we provide the dimensionality of each representation. Note that representations $\widecheck{\mathcal{O}}^{+}_{\{\Gamma, \Gamma'\}}$, $\overline{\mathcal{O}}^{\pm}_{\{\Gamma, \Gamma'\}}$, $\overline{\mathcal{O}}^{\pm}_{[\Gamma, \Gamma']}$ exist for $N_f>1$, representations $\widecheck{\mathcal{O}}^{\pm}_{[\Gamma, \Gamma']}$ for $N_f>2$, and  representation $\widecheck{\mathcal{O}}^{-}_{\{\Gamma, \Gamma'\}}$ for $N_f>3$.

\begin{figure}
\begin{eqnarray*}
\ytableausetup
{mathmode, boxframe=0.075em, boxsize=1.1em}
&& \quad N_f{-}1 \left\{
\begin{matrix}
\begin{ytableau}
\phantom{1} \\
 \\
\none[\svdots] \\
 \\
 \\
\end{ytableau}
\end{matrix} \right. \quad \bigotimes \quad 
N_f{-}1 \left\{
\begin{matrix}
\begin{ytableau}
\phantom{1} \\
 \\
\none[\svdots] \\
 \\
 \\
\end{ytableau} 
\end{matrix} \right. \quad \bigotimes \quad
\begin{matrix}
\begin{ytableau}
\phantom{1}
\end{ytableau} 
\end{matrix} \quad \bigotimes \quad
\begin{matrix}
\begin{ytableau}
\phantom{1}
\end{ytableau} 
\end{matrix} \quad \mathlarger{\mathlarger{\mathlarger{=}}} \quad \\
\\
\\
&& \left( 
N_f{-}1 \left\{
\begin{matrix}
\begin{ytableau}
\phantom{1} & \\
 & \\
\none[\svdots] & \none[\svdots] \\
 & \\
 & \\
\end{ytableau}
\end{matrix} \right. \quad \bigoplus \quad 
N_f \left\{
\begin{matrix}
\begin{ytableau}
\phantom{1} & \\
 & \\
\none[\svdots] & \none[\svdots] \\
 & \\
 \\
 \\
\end{ytableau}
\end{matrix} 
\right. \right) \quad \bigotimes \quad
\left( \begin{matrix}
\begin{ytableau}
\phantom{1} & \\
\end{ytableau} 
\end{matrix} \quad \bigoplus \quad
\begin{matrix}
\begin{ytableau}
\phantom{1} \\
 \\
\end{ytableau} 
\end{matrix} \right) \quad \mathlarger{\mathlarger{\mathlarger{=}}} \quad \end{eqnarray*}

\begin{eqnarray*}
&& N_f{-}1 \left\{
\begin{matrix}
\begin{ytableau}
\phantom{1} & & & \\
 & \\
\none[\svdots] & \none[\svdots] \\
 & \\
 & \\
\end{ytableau}
\end{matrix} \right. \quad \bigoplus \quad 
N_f \left\{
\begin{matrix}
\begin{ytableau}
\phantom{1} & & \\
 & \\
\none[\svdots] & \none[\svdots] \\
 & \\
 & \\ 
 \\
\end{ytableau} 
\end{matrix} \right. \quad \bigoplus \quad
N_f \left\{
\begin{matrix}
\begin{ytableau}
\phantom{1} & \\
 & \\
\none[\svdots] & \none[\svdots] \\
 & \\
 & \\
 & \\
\end{ytableau} 
\end{matrix} \right. \quad \bigoplus \quad
N_f{-}1 \left\{
\begin{matrix}
\begin{ytableau}
\phantom{1} & & \\
 & & \\
\none[\svdots] & \none[\svdots] \\
 & \\
 & \\
\end{ytableau} 
\end{matrix} \right. \quad \bigoplus \quad
N_f \left\{
\begin{matrix}
\begin{ytableau}
\phantom{1} & & \\
 & \\
\none[\svdots] & \none[\svdots] \\
 & \\
 & \\ 
 \\
\end{ytableau}
\end{matrix} \right. \quad \bigoplus \\
&& \\
&& \qquad \ (\widecheck{\mathcal{O}}^{+}_{\{\Gamma, \Gamma'\}})^{f_1, f_2}_{f_3, f_4} \qquad \qquad \ (\overline{\mathcal{O}}^{+}_{\{\Gamma, \Gamma'\}})^{f_1, f_2}_{f_3, f_4} \qquad \qquad (\widehat{\mathcal{O}}^{+}_{\{\Gamma, \Gamma'\}})^{f_1, f_2}_{f_3, f_4} \qquad \qquad \ (\widecheck{\mathcal{O}}^{-}_{[\Gamma, \Gamma']})^{f_1, f_2}_{f_3, f_4} \qquad \qquad \ (\widehat{\mathcal{O}}^{-}_{[\Gamma, \Gamma']})^{f_1, f_2}_{f_3, f_4} \\
\\
\\
&& \quad N_f \left\{
\begin{matrix}
\begin{ytableau}
\phantom{1} & & & \\
 & \\
\none[\svdots] & \none[\svdots] \\
 & \\
 \\
 \\
\end{ytableau} 
\end{matrix} \right. \quad \bigoplus \quad
N_f \left\{
\begin{matrix}
\begin{ytableau}
\phantom{1} & & \\
 & \\
\none[\svdots] & \none[\svdots] \\
 & \\
 & \\
 \\
\end{ytableau}
\end{matrix} \right. \quad \bigoplus \quad
N_f \left\{
\begin{matrix} 
\begin{ytableau}
\phantom{1} & & \\
 & & \\
\none[\svdots] & \none[\svdots] \\
 & \\
 \\
 \\
\end{ytableau} 
\end{matrix} \right. \quad \bigoplus \quad
N_f \left\{
\begin{matrix} 
\begin{ytableau}
\phantom{1} & & \\
 & \\
\none[\svdots] & \none[\svdots] \\
 & \\
 & \\
 \\
\end{ytableau} 
\end{matrix} \right. \quad \bigoplus \quad
N_f \left\{
\begin{matrix} 
\begin{ytableau}
\phantom{1} & \\
 & \\
\none[\svdots] & \none[\svdots] \\
 & \\
 & \\
 & \\
\end{ytableau}
\end{matrix} \right. \\
&& \\
&& \qquad (\widecheck{\mathcal{O}}^{+}_{[\Gamma, \Gamma']})^{f_1, f_2}_{f_3, f_4} \qquad \qquad \ 
(\widehat{\mathcal{O}}^{+}_{[\Gamma, \Gamma']})^{f_1, f_2}_{f_3, f_4} \qquad \qquad \quad (\widecheck{\mathcal{O}}^{-}_{\{\Gamma, \Gamma'\}})^{f_1, f_2}_{f_3, f_4} \qquad \qquad \ (\overline{\mathcal{O}}^{-}_{\{\Gamma, \Gamma'\}})^{f_1, f_2}_{f_3, f_4} \qquad \qquad (\widehat{\mathcal{O}}^{-}_{\{\Gamma, \Gamma'\}})^{f_1, f_2}_{f_3, f_4} 
\end{eqnarray*}
\caption{Young tableaux of four-quark operators for $SU(N_f)$ group.}
\label{Young_tableau}
\end{figure}

Operators that belong to different sets cannot mix among themselves under renormalization. In addition, operators $\overline{\mathcal{O}}^{\pm}$ and $\widehat{\mathcal{O}}^{\pm}$ can mix with lower-dimensional operators. The latter may contain at most one quark-antiquark pair, and thus, under the flavor group they can transform only under representations stemming from the tensor product of the fundamental and anti-fundamental representations, i.e. the singlet and the adjoint representations. As seen in Fig.~\ref{Young_tableau}, operators $\overline{\mathcal{O}}^{\pm}$ and $\widehat{\mathcal{O}}^{\pm}$ support precisely these representations, thus allowing them to mix with lower-dimensional operators. On the contrary, operators $\widecheck{\mathcal{O}}^{\pm}$ support representations that are distinct from the above. In the present work, we limit our study to operators $\widecheck{\mathcal{O}}^{\pm}$, which are free of mixing with lower-dimensional operators. The renormalization of operators $\overline{\mathcal{O}}^{\pm}$ and $\widehat{\mathcal{O}}^{\pm}$ will be addressed in a follow-up work.

\renewcommand{\arraystretch}{1.5}
\begin{table}[ht!]
    \centering
    \begin{tabular}{c|c|c}
    Operator & Type & \ Dimensionality \\
         \hline
           \hline
$(\widecheck{\mathcal{O}}^{\pm}_{\{\Gamma, \Gamma'\}})^{f_1, f_2}_{f_3, f_4}$ \ & \ traceless \ & $N_f^2 (N_f \pm 3) (N_f \mp 1)/4$ \\[2ex]
\hline
$(\overline{\mathcal{O}}^{\pm}_{\{\Gamma, \Gamma'\}})^{f_1, f_2}_{f_3, f_4}$ \ & \ adjoint \ & $N_f^2 - 1$ \\[2ex]
\hline
$(\widehat{\mathcal{O}}^{\pm}_{\{\Gamma, \Gamma'\}})^{f_1, f_2}_{f_3, f_4}$ \ & \ singlet \ & $1$ \\[2ex]
\hline
$(\widecheck{\mathcal{O}}^{\pm}_{[\Gamma, \Gamma']})^{f_1, f_2}_{f_3, f_4}$ \ & \ traceless \ & $(N_f^2 - 4) (N_f^2 - 1)/4$ \\[2ex]
\hline
$(\overline{\mathcal{O}}^{\pm}_{[\Gamma, \Gamma']})^{f_1, f_2}_{f_3, f_4}$ \ & \ adjoint \ & $N_f^2 - 1$
        \end{tabular}
    \caption{The dimensionality of the 10 irreducible representations of SU($N_f$) flavor group for the four-quark operators.}
\label{tab:dimensionalities}
\end{table}
\renewcommand{\arraystretch}{1}

\subsection{Scalar and pseudoscalar four-quark operators}
\label{sec:(pseudo)scalar}

Given that we are looking for operators which will be introduced as ingredients to an effective Lagrangian, we will focus exclusively on scalar ($\Gamma' = \Gamma$) and pseudoscalar ($\Gamma' = \Gamma \gamma_5$) four-quark operators. Further, as explained above, requiring absence of mixing with lower-dimensional operators leads us to the operators $\widecheck{\mathcal{O}}^{\pm}_{\{\Gamma, \Gamma'\}}$ and $\widecheck{\mathcal{O}}^{\pm}_{[\Gamma, \Gamma']}$. Due to different properties under parity (see section \ref{other_symmetries}), scalar and pesudoscalar operators are classified into different mixing sets. Thus, in this case, the operator basis contains 6 sets of operators (2 for scalar, and 4 for pseudoscalar):
\begin{eqnarray}
{S}_1^{\pm} &:& \Big\{ Q_1^\pm, \ Q_2^\pm, \ Q_3^\pm, \ Q_4^\pm, \ Q_5^\pm \Big\}, \label{S1} \\
{S}_2^{\pm} &:& \Big\{ \mathcal{Q}_1^\pm, \ \mathcal{Q}_4^\pm, \ \mathcal{Q}_5^\pm \Big\}, \label{S2} \\
{S}_3^{\pm} &:& \Big\{\mathcal{Q}_2^\pm, \ \mathcal{Q}_3^\pm \Big\}, \label{S3} 
\end{eqnarray}
where
\begin{equation}
\begin{split}
Q^\pm_1 &\equiv \widecheck{{\cal O}}_{VV+AA}^\pm, \\
Q^\pm_2 &\equiv \widecheck{{\cal O}}_{VV-AA}^\pm, \\
Q^\pm_3 &\equiv \widecheck{{\cal O}}_{SS-PP}^\pm, \\
Q^\pm_4 &\equiv \widecheck{{\cal O}}_{SS+PP}^\pm, \\
Q^\pm_5 &\equiv \widecheck{{\cal O}}_{TT}^\pm,
\end{split}
\qquad \qquad 
\begin{split}
\mathcal{Q}^\pm_1 &\equiv \widecheck{{\cal O}}_{VA+AV}^\pm, \\
\mathcal{Q}^\pm_2 &\equiv \widecheck{{\cal O}}_{VA-AV}^\pm, \\
\mathcal{Q}^\pm_3 &\equiv \widecheck{{\cal O}}_{PS-SP}^\pm, \\
\mathcal{Q}^\pm_4 &\equiv \widecheck{{\cal O}}_{PS+SP}^\pm, \\
\mathcal{Q}^\pm_5 &\equiv \widecheck{{\cal O}}_{T \tilde{T}}^\pm. 
\end{split}
\label{Q_definitions}
\end{equation}

\noindent Note that $\widecheck{\mathcal O}_{\Gamma\Gamma \pm \Gamma'\Gamma'}$ stands for $(\widecheck{\mathcal O}_{\{\Gamma, \Gamma\}} \pm \widecheck{\mathcal O}_{\{\Gamma', \Gamma' \}})/2$, and
$\widecheck{\mathcal O}_{T\Tilde{T}}$ stands for $\widecheck{\mathcal O}_{\{T, \tilde{T}\}}/2$. The operators in the sets $S_1^\pm$ are scalar and originate from the representations $\widecheck{\mathcal{O}}_{\{\Gamma,\Gamma\}}^\pm$.\footnote{Specific combinations of operators supporting the same representation have been taken, which, as stated in the sequel, are eigenstates of chiral transformations and are more useful in phenomenology.} The operators in the sets $S_2^\pm$ and $S_3^\pm$ are pseudoscalar, and they stem from the representations $\widecheck{\mathcal{O}}_{\{\Gamma,\Gamma \gamma_5\}}^\pm$ and $\widecheck{\mathcal{O}}_{[\Gamma,\Gamma \gamma_5]}^\pm$, respectively. The operators $\widecheck{{\cal O}}_{\tilde T T}^\pm$ and $\widecheck{{\cal O}}_{\tilde T \tilde T}^\pm$ are not explicitly shown above, as they coincide with $\widecheck{{\cal O}}_{T \tilde T}^\pm$ and $\widecheck{{\cal O}}_{T T}^\pm$, respectively. Also, a summation over all independent Lorentz indices (if any) of the Dirac matrices is understood. In particular, $\widecheck{{\cal O}}_{T T}^\pm$, involving two Lorentz indices, contains a sum of 12 (pairwise equal) contributions. The notation used above ($Q_i^\pm$, $\mathcal{Q}_i^\pm$) is consistent with older studies \cite{Donini:1999sf,Papinutto:2016xpq} for $\Delta F = 2$ operators.\footnote{In some papers, e.g., Ref.\cite{Papinutto:2016xpq}, operators $Q_5$ and ${\cal Q}_5$ have an additional factor of 2.} In the case of $\Delta F<2$ operators, the relevant trace parts are subtracted (they are identically zero for $\Delta F = 2$).   

We stress that operators belonging to the same set can mix among themselves because they support the same representation of the flavor group. For the same reason (combined with invariance under parity), mixing is absent among operators belonging to different sets. As we conclude in the next subsection, operator sets $S_2^\pm$ are further decomposed into two individual subsets, and thus mixing can exist only within each subset.

The operator basis can be also expressed in terms of Fierz-transformed operators given in Eqs. (\ref{fourQFD},  \ref{fourQFDF}). Using the Fierz rearrangement identity for Euclidean gamma matrices in 4 dimensions and for Grassmann fields: 
\begin{eqnarray}
    && \Big(\sum_{a} \sum_{\alpha,\beta} {\bar \psi}_{\alpha}^{a, f_1}(x)  \,\Gamma^A_{\alpha \beta} \,  \psi_{\beta, f_3}^{a}(x) \Big) \Big(\sum_{a'} \sum_{\alpha',\beta'} {\bar \psi}_{\alpha'}^{a',f_2}(x) \,\Gamma^B_{\alpha' \beta'} \, \psi_{\beta',f_4}^{a'}(x) \Big) = \nonumber \\
    && \sum_{C,D} c^{AB}_{CD} \Big[\sum_{a,a'} \Big(\sum_{\alpha,\beta} {\bar \psi}_{\alpha}^{a, f_1}(x)  \,\Gamma^C_{\alpha \beta} \,  \psi_{\beta, f_4}^{a'}(x) \Big) \Big(\sum_{\alpha',\beta'} {\bar \psi}_{\alpha'}^{a',f_2}(x) \,\Gamma^D_{\alpha' \beta'} \, \psi_{\beta',f_3}^{a}(x) \Big)\Big], 
\end{eqnarray}
where $A,\, B,\, C,\, D = 1, 2,\ldots, 16$ run over all 16 independent $\gamma$-matrices in 4 dimensions and
\begin{equation}
    c^{AB}_{CD} = - \frac{1}{16} {\rm tr} \left[\Gamma^A (\Gamma^D)^\dagger \Gamma^B (\Gamma^C)^\dagger \right], 
\end{equation}
we extract the following relations between the original and Fierz-transformed operators:
\begin{equation}
\hspace{-0.8cm}\begin{pmatrix*}[r]
{\mathcal{O}_{VV}} \\[0.4em]
{\mathcal{O}_{AA}} \\[0.4em]
{\mathcal{O}_{SS}} \\[0.4em]
{\mathcal{O}_{PP}} \\[0.4em]
{\mathcal{O}_{TT}}
\end{pmatrix*} = \frac{1}{8} \begin{pmatrix*}[r]
4 & 4 & -8 & 8 & \phantom{+}0 \\[0.4em]
4 & 4 & 8 & -8 & 0 \\[0.4em]
-2 & 2 & -2 & -2 & 1 \\[0.4em]
2 & -2 & -2 & -2 & 1 \\[0.4em]
0 & 0 & 24 & 24 & 4 
\end{pmatrix*} \begin{pmatrix*}[r]
{\mathcal{O}^{F_D}_{VV}}\\[0.4em]
{\mathcal{O}^{F_D}_{AA}} \\[0.4em]
{\mathcal{O}^{F_D}_{SS}} \\[0.4em] 
{\mathcal{O}^{F_D}_{PP}} \\[0.4em]
{\mathcal{O}^{F_D}_{TT}}
\end{pmatrix*}, \qquad 
\begin{pmatrix*}[r]
{\mathcal{O}_{VA}} \\[0.4em]
{\mathcal{O}_{AV}} \\[0.4em]
{\mathcal{O}_{PS}} \\[0.4em]
{\mathcal{O}_{SP}} \\[0.4em]
{\mathcal{O}_{T\tilde{T}}}
\end{pmatrix*} =\frac{1}{8} \begin{pmatrix*}[r]
4 & 4 & -8 & 8 & \phantom{+}0 \\[0.4em]
4 & 4 & 8 & -8 & 0 \\[0.4em]
-2 & 2 & -2 & -2 & 1 \\[0.4em]
2 & -2 & -2 & -2 & 1 \\[0.4em]
0 & 0 & 24 & 24 & 4 
\end{pmatrix*} \begin{pmatrix*}[r]
{\mathcal{O}^{F_D}_{VA}}\\[0.4em]
{\mathcal{O}^{F_D}_{AV}} \\[0.4em]
{\mathcal{O}^{F_D}_{PS}} \\[0.4em] 
{\mathcal{O}^{F_D}_{SP}} \\[0.4em]
{\mathcal{O}^{F_D}_{T\tilde{T}}}
\end{pmatrix*}.
\end{equation}
It is worth noting that the transformation matrix is the same between the scalar and pseudoscalar operators. Our results agree with Ref.~\cite{Donini:1999sf}. It is straightforward to construct the transformation matrix for the $Q_i^\pm$ and $\mathcal{Q}_i^\pm$ operators:
\begin{equation}
\begin{pmatrix*}[r]
Q_1^\pm \\[0.4em]
Q_2^\pm \\[0.4em]
Q_3^\pm \\[0.4em]
Q_4^\pm \\[0.4em]
Q_5^\pm
\end{pmatrix*} = \frac{1}{4} \begin{pmatrix*}[r]
\phantom{+}4 & 0 & 0 & 0 & \phantom{+}0 \\[0.4em]
0 & 0 & -8 & 0 & 0 \\[0.4em]
0 & -2 & 0 & 0 & 0 \\[0.4em]
0 & 0 & 0 & -2 & 1 \\[0.4em]
0 & 0 & 0 & 12 & 2 
\end{pmatrix*} \begin{pmatrix*}[r]
(Q_1^\pm)^{F_D} \\[0.4em]
(Q_2^\pm)^{F_D} \\[0.4em]
(Q_3^\pm)^{F_D} \\[0.4em] 
(Q_4^\pm)^{F_D} \\[0.4em]
(Q_5^\pm)^{F_D}
\end{pmatrix*}, \qquad 
\begin{pmatrix*}[r]
\mathcal{Q}_1^\pm \\[0.4em]
\mathcal{Q}_2^\pm \\[0.4em]
\mathcal{Q}_3^\pm \\[0.4em]
\mathcal{Q}_4^\pm \\[0.4em]
\mathcal{Q}_5^\pm
\end{pmatrix*} = \frac{1}{4} \begin{pmatrix*}[r]
\phantom{+}4 & 0 & 0 & 0 & \phantom{+}0 \\[0.4em]
0 & 0 & -8 & 0 & 0 \\[0.4em]
0 & -2 & 0 & 0 & 0 \\[0.4em]
0 & 0 & 0 & -2 & 1 \\[0.4em]
0 & 0 & 0 & 12 & 2 
\end{pmatrix*} \begin{pmatrix*}[r]
(\mathcal{Q}_1^\pm)^{F_D} \\[0.4em]
(\mathcal{Q}_2^\pm)^{F_D} \\[0.4em]
(\mathcal{Q}_3^\pm)^{F_D} \\[0.4em] 
(\mathcal{Q}_4^\pm)^{F_D} \\[0.4em]
(\mathcal{Q}_5^\pm)^{F_D}
\end{pmatrix*}.
\label{QtoQFD}
\end{equation}
The block-diagonal form of the transformation matrices in Eq.~(\ref{QtoQFD}) agrees with the expected form of the mixing matrices (i.e., a Fierz-transformed operator cannot be a linear combination of operators supporting different representations of the flavor group), as demonstrated in the next subsection by studying other symmetries of the action. 

\subsection{Other symmetry properties}
\label{other_symmetries}
In addition to flavor symmetry, other symmetry transformations can be used to decompose the mixing pattern into smaller sets. We examine the action of parity $\mathcal{P}$ and charge conjugation $\mathcal{C}$ transformations on each scalar and pseudoscalar operator of the basis given in Eqs. (\ref{S1} -- \ref{S3}. We first define the $\mathcal{P}$ and $\mathcal{C}$ transformations on quark and antiquark fields:
\begin{eqnarray}
\rm{Parity:}
&&\begin{cases}
    \mathcal{P}\psi_f(x) & = \gamma_4 \ \psi_f(x_P) \\
    \mathcal{P}\bar{\psi}_f(x) &= \bar{\psi}_f(x_P) \ \gamma_4,
\end{cases} \\
\rm{Charge\,conjugation:}
&&\begin{cases}
    \mathcal{C}\psi_f(x) &= C^{-1} \ \bar{\psi}_f^T(x) \\
    \mathcal{C}\bar{\psi}_f(x) &= -\psi_f^T(x) \ C,
\end{cases}
\end{eqnarray}
where $x_P = (-\vec{x},t)$, and $C \gamma_{\mu} C^{-1} = - \gamma_{\mu}^T$. In Table \ref{tb1} we list the symmetry transformations for each operator.
\renewcommand{\arraystretch}{1.5}
\begin{table}[ht!]
    \centering
    \begin{tabular}{c|c|c||c|c|c}
    Scalar & & & Pseudoscalar & & \\
     Operator &  \, \,$\mathcal{P}$ (about $\vec{x} = 0$) \,\,& \, \,$\mathcal{C}$\,\, & Operator &  \, \,$\mathcal{P}$ (about $\vec{x} = 0$) \,\,& \, \,$\mathcal{C}$\,\,\\
         \hline
           \hline
    $(Q_1^{\pm})^{f_1, f_2}_{f_3,f_4}$ \, \, & $+(Q_1^{\pm})^{f_1, f_2}_{f_3,f_4}$ \, \, & $+(Q_1^{\pm})^{f_3, f_4}_{f_1,f_2}$ \, \, & $(\mathcal{Q}_1^{\pm})^{f_1, f_2}_{f_3,f_4}$ \, \, & $-(\mathcal{Q}_1^{\pm})^{f_1, f_2}_{f_3,f_4}$ \, \, & $-(\mathcal{Q}_1^{\pm})^{f_3, f_4}_{f_1,f_2}$ \, \, \\ [1em] 
    \hline
    $(Q_2^{\pm})^{f_1, f_2}_{f_3,f_4}$ \, \, & $+(Q_2^{\pm})^{f_1, f_2}_{f_3,f_4}$ \, \, & $+(Q_2^{\pm})^{f_3, f_4}_{f_1,f_2}$ \, \, & $(\mathcal{Q}_2^{\pm})^{f_1, f_2}_{f_3,f_4}$ \, \, & $-(\mathcal{Q}_2^{\pm})^{f_1, f_2}_{f_3,f_4}$ \, \, & $-(\mathcal{Q}_2^{\mp})^{f_3, f_4}_{f_1,f_2}$ \, \, \\ [1em]  
    \hline
    $(Q_3^{\pm})^{f_1, f_2}_{f_3,f_4}$ \, \, & $+(Q_3^{\pm})^{f_1, f_2}_{f_3,f_4}$ \, \, & $+(Q_3^{\pm})^{f_3, f_4}_{f_1,f_2}$ \, \, & $(\mathcal{Q}_3^{\pm})^{f_1, f_2}_{f_3,f_4}$ \, \, & $-(\mathcal{Q}_3^{\pm})^{f_1, f_2}_{f_3,f_4}$ \, \, & $+(\mathcal{Q}_3^{\mp})^{f_3, f_4}_{f_1,f_2}$ \, \, \\ [1em]  
    \hline
    $(Q_4^{\pm})^{f_1, f_2}_{f_3,f_4}$ \, \, & $+(Q_4^{\pm})^{f_1, f_2}_{f_3,f_4}$ \, \, & $+(Q_4^{\pm})^{f_3, f_4}_{f_1,f_2}$ \, \, & $(\mathcal{Q}_4^{\pm})^{f_1, f_2}_{f_3,f_4}$ \, \, & $-(\mathcal{Q}_4^{\pm})^{f_1, f_2}_{f_3,f_4}$ \, \, & $+(\mathcal{Q}_4^{\pm})^{f_3, f_4}_{f_1,f_2}$ \, \, \\ [1em]  
    \hline
    $(Q_5^{\pm})^{f_1, f_2}_{f_3,f_4}$ \, \, & $+(Q_5^{\pm})^{f_1, f_2}_{f_3,f_4}$ \, \, & $+(Q_5^{\pm})^{f_3, f_4}_{f_1,f_2}$ \, \, & $(\mathcal{Q}_5^{\pm})^{f_1, f_2}_{f_3,f_4}$ \, \, & $-(\mathcal{Q}_5^{\pm})^{f_1, f_2}_{f_3,f_4}$ \, \, & $+(\mathcal{Q}_5^{\pm})^{f_3, f_4}_{f_1,f_2}$ \, \, \\ [1em]    
        \end{tabular}
    \caption{Transformations of the four-quark operators under $\mathcal{P}$ (considering operators localized at $\vec{x} =0$) 
    and $\mathcal{C}$.}
\label{tb1}
\end{table}
\renewcommand{\arraystretch}{1}

The scalar (pseudoscalar) operators are even (odd) under parity, as expected, while none of the operators are eigenstates of charge conjugation. To address this, one can change the basis by considering combinations of operators in which flavor indices of quark and antiquark fields are swapped, specifically $(f_1 \leftrightarrow f_3 , f_2 \leftrightarrow f_4)$. This operation mixes different components within the same set of operators supporting a given representation of the flavor group, and thus it only mixes objects which will be renormalized in the same way. 

For the scalar operators $Q_i^\pm$, the construction of eigenstates of $\mathcal{C}$ does not affect the mixing pattern and thus it is inconsequential to the study of operator renormalization. Therefore, we maintain the original basis of scalar operators given in Eq. \eqref{S1}.

For operators $\mathcal{Q}_2^\pm$ and $\mathcal{Q}_3^\pm$, flavor swapping by itself does not produce eigenstates of charge conjugation, as plus operators transform into minus operators under $\mathcal{C}$ and vice versa. Although it is possible to construct eigenstates of $\mathcal{C}$ by combining plus and minus operators, these combinations merge operators that support different representations of the flavor group. This reflects the fact that, in this case, charge conjugation does not commute with flavor transformations; thus, there are no simultaneous eigenstates of charge conjugation and flavor transformations. In such a case, we need to select between the construction of eigenstates of $\mathcal{C}$ or flavor transformations. Following older studies, we keep the form of $\mathcal{Q}_2^\pm$ and $\mathcal{Q}_3^\pm$, which are classified according to the flavor symmetry and not $\mathcal{C}$. However, $\mathcal{C}$ can be used to extract relations between the mixing coefficients of the corresponding plus and minus operators, as discussed in Section \ref{Mixing_matrices}.  

For the operators $\mathcal{Q}_1^\pm$, $\mathcal{Q}_4^\pm$, and $\mathcal{Q}_5^\pm$, we construct eigenstates of $\mathcal{C}$, as follows: 
\begin{eqnarray}
    && \Big\{(\mathcal{Q}_1^\pm)^{f_1, f_2}_{f_3,f_4} - (\mathcal{Q}_1^{\pm})^{f_3, f_4}_{f_1,f_2}, \ 
    (\mathcal{Q}_4^\pm)^{f_1, f_2}_{f_3,f_4} + (\mathcal{Q}_4^{\pm})^{f_3, f_4}_{f_1,f_2}, \
    (\mathcal{Q}_5^\pm)^{f_1, f_2}_{f_3,f_4} + (\mathcal{Q}_5^{\pm})^{f_3, f_4}_{f_1,f_2} \Big\}, \label{S2_a} \\
&& \Big\{(\mathcal{Q}_1^\pm)^{f_1, f_2}_{f_3,f_4} + (\mathcal{Q}_1^{\pm})^{f_3, f_4}_{f_1,f_2}, \ 
    (\mathcal{Q}_4^\pm)^{f_1, f_2}_{f_3,f_4} - (\mathcal{Q}_4^{\pm})^{f_3, f_4}_{f_1,f_2}, \
    (\mathcal{Q}_5^\pm)^{f_1, f_2}_{f_3,f_4} - (\mathcal{Q}_5^{\pm})^{f_3, f_4}_{f_1,f_2} \Big\}. \label{S2_b} 
\end{eqnarray}
Operators in Eq. \eqref{S2_a} [Eq. \eqref{S2_b}] are even (odd) under $\mathcal{C}$. Thus, the two sets are renormalized separately through a common $3 \times 3$ mixing matrix, since they still support the same representation of the flavor group. The mixing matrix is independent of the flavor content of the operators. For particular choices of the flavor indices $f_1$, $f_2$, $f_3$ and $f_4$, the first operator in Eq. \eqref{S2_a} and the last two operators in Eq. \eqref{S2_b} vanish simultaneously. This indicates that this $3 \times 3$ matrix actually has to be block diagonal of the form: $(1 \times 1) \oplus (2 \times 2)$. Then, operator $\mathcal{Q}_1^\pm$ renormalizes independently of $(\mathcal{Q}_4^\pm, \mathcal{Q}_5^\pm)$ and the mixing sets $S_2^\pm$ are decomposed into two smaller subsets:
\begin{eqnarray}
  {S}_{2a}^{\pm}&:& \Big\{ \mathcal{Q}_1^\pm \Big\}, \\
  {S}_{2b}^{\pm}&:& \Big\{ \mathcal{Q}_4^\pm, \mathcal{Q}_5^\pm \Big\}.
\end{eqnarray}
Here, we keep the original form of $\mathcal{Q}_1^\pm$, $\mathcal{Q}_4^\pm$, and $\mathcal{Q}_5^\pm$, as given in Eq. \eqref{S2}. Ultimately, the construction of eigenstates of $\mathcal{C}$ serves only to show that the mixing sets $S_2^\pm$ can be further broken down, as well as to show relations between the mixing coefficients of the operators in $S_3^+$ and $S_3^-$ (see Sec. \ref{Mixing_matrices}).

For chirality preserving actions, the mixing sets $S_1^\pm$ can be further decomposed into three sets. We define two types of axial chiral transformations:
\begin{eqnarray}
\rm{Chiral\,transformations\,1\ }
&\bigl({\rm U}(1)_\mathcal{A}\bigr)\phantom{S_f}\ :&\begin{cases}
    \mathcal{A}_1 \psi_f(x) & = e^{i \alpha \gamma_5} \psi_{f} (x) \\
    \mathcal{A}_1 \bar{\psi}^f(x) &= \bar{\psi}^{f} (x) \ e^{i \alpha \gamma_5}
\end{cases}, \\
\rm{Chiral\,transformations\,2\ }
&\bigl({\rm SU}(N_f)_\mathcal{A}\bigr)\ :&\begin{cases}
    \mathcal{A}_2 \psi_f(x) &= \sum_{f'} \ \left(e^{i \alpha \gamma_5 T^a}\right)_f {}^{f'} \psi_{f'} (x) \\
    \mathcal{A}_2 \bar{\psi}^f(x) &= \sum_{f'} \ \bar{\psi}^{f'} (x) \ \left(e^{i \alpha \gamma_5 T^a}\right)_{f'} {}^f
\end{cases},
\end{eqnarray}
where $T^a$ is a generator of the $su(N_f)$ algebra. A four-quark operator with general gamma matrices transforms under the two chiral transformations as follows:
\begin{eqnarray}
    \mathcal{A}_1 (\mathcal{O}_{\Gamma \Gamma'})^{f_1,f_2}_{f_3,f_4} &=& \Big(\bar{\psi}^{f_1} e^{i \alpha \gamma_5} \ \Gamma \ e^{i \alpha \gamma_5} \psi_{f_3} \Big) \ \Big(\bar{\psi}^{f_2} e^{i \alpha \gamma_5} \ \Gamma' \ e^{i \alpha \gamma_5} \psi_{f_4} \Big), \\
    \mathcal{A}_2 (\mathcal{O}_{\Gamma \Gamma'})^{f_1,f_2}_{f_3,f_4} &=& \left( \sum_{f,f'}\bar{\psi}^{f} \left(e^{i \alpha \gamma_5 T^a}\right)_f {}^{f_1} \ \Gamma \left(e^{i \alpha \gamma_5 T^a}\right)_{f_3} {}^{f'} \psi_{f'} \right) \ \left( \sum_{f,f'}\bar{\psi}^{f} \left(e^{i \alpha \gamma_5 T^a}\right)_f {}^{f_2} \ \Gamma' \left(e^{i \alpha \gamma_5 T^a}\right)_{f_4} {}^{f'} \psi_{f'} \right), 
\end{eqnarray}
where 
\begin{eqnarray}
    e^{i \alpha \gamma_5} \ \Gamma \ e^{i \alpha \gamma_5} &=& \begin{cases}
        \Gamma, & \qquad \qquad \qquad \qquad \quad \ \ \, \{ \gamma_5, \Gamma \} = 0 \\
        \Gamma e^{2i \alpha \gamma_5}, & \qquad \qquad \qquad \qquad \quad \ \ \ [\gamma_5, \Gamma] = 0
    \end{cases}, \\
    \left(e^{i \alpha \gamma_5 T^a}\right)_{f} {}^{f_1} \ \Gamma \ \left(e^{i \alpha \gamma_5 T^a}\right)_{f_3} {}^{f'} &=& \begin{cases}
        \Gamma \left(e^{-i \alpha \gamma_5 T^a}\right)_{f} {}^{f_1} \left(e^{i \alpha \gamma_5 T^a}\right)_{f_3} {}^{f'}, & \{ \gamma_5, \Gamma \} = 0 \\
        \Gamma \left(e^{i \alpha \gamma_5 T^a}\right)_{f} {}^{f_1} \left(e^{i \alpha \gamma_5 T^a}\right)_{f_3} {}^{f'}, & \, [\gamma_5, \Gamma] = 0
    \end{cases}. \label{A2_transf}
\end{eqnarray}
By selecting a diagonal generator $(T^a)_{f_1} {}^{f_2} = \lambda_{f_1} \delta_{f_1} {}^{f_2}$, where $\sum_{i=1}^{N_f} \lambda_i = 0$, and $\lambda_i \in \mathbb{R}$, Eq. \eqref{A2_transf} can be simplified by using the following relation:
\begin{equation}
    \left(e^{\pm i \alpha \gamma_5 T^a}\right)_{f} {}^{f_1} \left(e^{i \alpha \gamma_5 T^a}\right)_{f_3} {}^{f'} = \delta_{f} {}^{f_1} \delta_{f_3} {}^{f'} 
    e^{i \alpha (\pm \lambda_{f_1} + \lambda_{f_3}) \gamma_5}.
\end{equation}
Then one can show that the operators or combinations of operators listed in Table \ref{tab:chiral_eigenstates} are eigenstates of $\mathcal{A}_1$ or $\mathcal{A}_2$ (when $T^a$ is diagonal). Eigenstates with different eigenvalues cannot mix. Using $\mathcal{A}_1$, we conclude that the operators $\{ Q_1^\pm, Q_2^\pm, Q_3^\pm \}$ are renormalized separately from $\{Q_4^\pm, Q_5^\pm\}$. Furthermore, using $\mathcal{A}_2$ with a diagonal $T^a$ and $\lambda_{f_1} \neq (\lambda_{f_3}, \lambda_{f_4})$, $\lambda_{f_2} \neq (\lambda_{f_3}, \lambda_{f_4})$, we conclude that operators $Q_1^\pm$ are also renormalized independently of $\{Q_2^\pm, \ Q_3^\pm \}$. The conclusion that $(Q_4, Q_5)$ do not mix with $(Q_1, Q_2, Q_3)$ for any $\lambda_i$ follows also from the $\mathcal{A}_2$ symmetry, without having to rely on the anomalous $\mathcal{A}_1$ symmetry. Thus, the $5 \times 5$ mixing matrices for the scalar operators of the sets $S_1^\pm$ are block diagonal of the form: $(1 \times 1) \oplus (2 \times 2) \oplus (2 \times 2)$, similar to the case of pseudoscalar operators. The three subsets of $S_1^\pm$ are summarized below: 
\begin{eqnarray}
    S_{1a}^\pm &:& \Big\{Q_1^\pm \Big\}, \\ 
    S_{1b}^\pm &:& \Big\{Q_2^\pm, \ Q_3^\pm \Big\}, \\
    S_{1c}^\pm &:& \Big\{Q_4^\pm, \ Q_5^\pm \Big\}.
\end{eqnarray} 
The eigenstates of the chiral transformations given in Table \ref{tab:chiral_eigenstates} can be used to extract relations between the mixing coefficients of the scalar and pseudoscalar operators, as discussed in Section \ref{Mixing_matrices}. We mention that in the above analysis we have deduced the mixing pattern of the four-quark operators without appealing to flavor switching symmetries applied to operators with distinct flavor indices $f_1$, $f_2$, $f_3$, $f_4$, as was done in older papers.

Given that not all lattice actions are chirally invariant, we do not employ chiral symmetry for determining the mixing matrices. Thus, in what follows we consider the 8 mixing sets $S_1^\pm$, $S_{2a}^\pm$, $S_{2b}^\pm$ and $S_3^\pm$.  

\renewcommand{\arraystretch}{1.75}
\begin{table}[ht!]
    \centering
    \begin{tabular}{c|c||c|c}
    Eigenstate of $\mathcal{A}_1$ \ & \ Eigenvalue \ & Eigenstate of $\mathcal{A}_2$ & Eigenvalue \\
         \hline
           \hline
$Q_1^\pm$ & 1 & $Q_1^\pm + \mathcal{Q}_1^\pm$ & \ \ $e^{+i \alpha (-\lambda_{f_1} - \lambda_{f_2} + \lambda_{f_3} + \lambda_{f_4})}$ \\[1ex]
\hline
$\mathcal{Q}_1^\pm$ & 1 & $Q_1^\pm - \mathcal{Q}_1^\pm$ & \ \ $e^{-i \alpha (-\lambda_{f_1} - \lambda_{f_2} + \lambda_{f_3} + \lambda_{f_4})}$ \\[1ex]
\hline
$Q_2^\pm$ & 1 & \ ($Q_2^+ \pm \mathcal{Q}_2^+) + (Q_2^- \pm \mathcal{Q}_2^-)$ \ & \ \ $e^{\pm i \alpha (+\lambda_{f_1} - \lambda_{f_2} - \lambda_{f_3} + \lambda_{f_4})}$ \\[1ex]
\hline
$\mathcal{Q}_2^\pm$ & 1 & $(Q_2^+ \pm \mathcal{Q}_2^+) - (Q_2^- \pm \mathcal{Q}_2^-)$ & \ \ $e^{\pm i \alpha (+\lambda_{f_1} - \lambda_{f_2} + \lambda_{f_3} - \lambda_{f_4})}$ \\[1ex]
\hline
$Q_3^\pm$ & 1 & $(Q_3^+ \pm \mathcal{Q}_3^+) + (Q_3^- \pm \mathcal{Q}_3^-)$ & \ \ $e^{\pm i \alpha (+\lambda_{f_1} - \lambda_{f_2} + \lambda_{f_3} - \lambda_{f_4})}$ \\[1ex]
\hline
$\mathcal{Q}_3^\pm$ & 1 & $(Q_3^+ \pm \mathcal{Q}_3^+) - (Q_3^- \pm \mathcal{Q}_3^-)$ & \ \ $e^{\pm i \alpha (+\lambda_{f_1} - \lambda_{f_2} - \lambda_{f_3} + \lambda_{f_4})}$ \\[1ex]
\hline
$Q_4^\pm + \mathcal{Q}_4^\pm$ & $e^{+i 4 \alpha}$ & $Q_4^\pm + \mathcal{Q}_4^\pm$ & \ \ $e^{+i \alpha (+\lambda_{f_1} + \lambda_{f_2} + \lambda_{f_3} + \lambda_{f_4})}$ \\[1ex]
\hline
$Q_4^\pm - \mathcal{Q}_4^\pm$ & $e^{-i 4 \alpha}$ & $Q_4^\pm - \mathcal{Q}_4^\pm$ & \ \ $e^{-i \alpha (+\lambda_{f_1} + \lambda_{f_2} + \lambda_{f_3} + \lambda_{f_4})}$ \\[1ex]
\hline
$Q_5^\pm + \mathcal{Q}_5^\pm$ & $e^{+i 4 \alpha}$ & $Q_5^\pm + \mathcal{Q}_5^\pm$ & \ \ $e^{+i \alpha (+\lambda_{f_1} + \lambda_{f_2} + \lambda_{f_3} + \lambda_{f_4})}$ \\[1ex]
\hline
$Q_5^\pm - \mathcal{Q}_5^\pm$ & $e^{-i 4 \alpha}$ & $Q_5^\pm - \mathcal{Q}_5^\pm$ & \ \ $e^{-i \alpha (+\lambda_{f_1} + \lambda_{f_2} + \lambda_{f_3} + \lambda_{f_4})}$ \\[1ex]
\hline
        \end{tabular}
    \caption{Eigenstates of $\mathcal{A}_1$ and $\mathcal{A}_2$ chiral transformations with their corresponding eigenvalues.}
\label{tab:chiral_eigenstates}
\end{table}
\renewcommand{\arraystretch}{1}

\subsection{Operators for flavor-changing processes}
\label{sec:DF}
The operator basis formulated in the previous subsections for the traceless representations of the flavor group allows one to study three cases of four-quark operators, which are involved in flavor-changing processes. We remind the reader that the operators studied in this work support representations which cannot have mixing with lower dimensional operators; even in these representations we encounter not only operators with $\Delta F=2$, but also with $\Delta F=1$ and $\Delta F=0$. The three cases are:
\begin{itemize}
    \item \underline{{\bf Case 1:} $f_1 \notin \{f_3, f_4\} $ and $f_2 \notin \{f_3, f_4\}$.} This case contains operators with $\Delta F=2$, when ($f_1 = f_2$ or $f_3 = f_4$), and operators with $\Delta F=1$, when $(f_1, f_2, f_3, f_4)$ are all different. The traceless  operators of Eqs. (\ref{Op1}, \ref{Op4}) are simplified to: 
\begin{equation}
    \widecheck{\mathcal{O}}^{\pm}_{\{\Gamma, \Gamma'\}} = \mathcal{O}^{\pm}_{\{\Gamma, \Gamma'\}}, \qquad \widecheck{\mathcal{O}}^{\pm}_{[\Gamma, \Gamma']} = \mathcal{O}^{\pm}_{[\Gamma, \Gamma']},    
    \end{equation}
where the subtracted trace terms simply vanish. As a result, operators $Q_i^\pm$ and $\mathcal{Q}_i^\pm$ are reduced to the operators studied in our previous work\cite{Constantinou:2024wdb}. Note that not all representations have $\Delta F=2$ operators. In particular, $\mathcal{O}^-_{\{\Gamma,\Gamma'\}} = 0$ when $f_1=f_2$ or $f_3 = f_4$, $\mathcal{O}^+_{[\Gamma,\Gamma']} = 0$ when $f_1=f_2$, and $\mathcal{O}^-_{[\Gamma,\Gamma']} = 0$ when $f_3=f_4$. Consequently, $Q_i^-$, $\mathcal{Q}_1^-$, $\mathcal{Q}_4^-$ and $\mathcal{Q}_5^-$ are not $\Delta F = 2$ operators. Similarly, operators $\mathcal{Q}_2^+$, $\mathcal{Q}_3^+$ are $\Delta F=2$ only when $f_3=f_4$, and $\mathcal{Q}_2^-$, $\mathcal{Q}_3^-$ only when $f_1=f_2$.  
    \item \underline{{\bf Case 2:} ($f_1 \notin \{f_3, f_4\} $ and $f_2 \in \{f_3, f_4\}$) or ($f_1 \in \{f_3, f_4\} $ and $f_2 \notin \{f_3, f_4\}$).} This case contains operators with $\Delta F=1$. The traceless operators of Eqs. (\ref{Op1}, \ref{Op4}) are simplified to: 
\begin{equation}
    \widecheck{\mathcal{O}}^{\pm}_{\{\Gamma, \Gamma'\}} = \mathcal{O}^{\pm}_{\{\Gamma, \Gamma'\}} - \frac{1}{N_f \pm 2} \overline{\mathcal{O}}^\pm_{\{ \Gamma, \Gamma'\}}, \qquad \widecheck{\mathcal{O}}^{\pm}_{[\Gamma, \Gamma']} = \mathcal{O}^{\pm}_{\{\Gamma, \Gamma'\}} - \frac{1}{N_f} \overline{\mathcal{O}}^\pm_{[ \Gamma, \Gamma']},    
    \end{equation}
where the subtracted pure trace terms are eliminated, thus simplifying operators $Q_i^\pm$ and $\mathcal{Q}_i^\pm$ accordingly.
    \item \underline{{\bf Case 3:} $(f_1,f_2)$ are pairwise equal to $(f_3,f_4)$.} This case contains operators with $\Delta F=0$. There are no simplifications in the definition of the traceless operators of Eqs. (\ref{Op1}, \ref{Op4}); all terms survive.
\end{itemize}
We note that each of the four $\widecheck{\mathcal{O}}$ representations contains at least some operators belonging to Case 1, i.e., operators in which the Kronecker deltas multiplying trace contributions have a vanishing value. This means that operators supporting the same $\widecheck{\mathcal{O}}$ representation but having different values of $\Delta F$ share the same renormalization, since they transform under the same representation of the flavor group. Then it is natural to extract the mixing matrices using the simplest case of operators, i.e., Case 1, where all subtracted trace terms are zero. Thus, our analysis developed in Ref.~\cite{Constantinou:2024wdb} can be also applied to all traceless operators of $\Delta F \leq 2$. 

\section{Renormalization setup}
\label{renorm}

In the following, we establish the setup for the renormalization of the scalar and pseudoscalar four-quark operators $Q_i^\pm$ and $\mathcal{Q}_i^\pm$. We define the corresponding mixing matrices and specify the Green’s functions that must be computed to construct a gauge-invariant renormalization scheme.

\subsection{Mixing matrices}
\label{Mixing_matrices}

By considering the mixing sets $
\big\{ Q_1^\pm, \ Q_2^\pm, \ Q_3^\pm, \ Q_4^\pm, \ Q_5^\pm \big\},\ \big\{ \mathcal{Q}_1^\pm\big\},\ \big\{\mathcal{Q}_2^\pm, \ \mathcal{Q}_3^\pm \big\},\  \big\{ \mathcal{Q}_4^\pm, \ \mathcal{Q}_5^\pm \big\}, $ derived in the previous section, the $5 \times 5$ mixing matrices $Z^\pm$ (${\cal Z}^\pm$), which renormalize the scalar (pseudoscalar) operators, $Q_i^\pm$ ($\mathcal{Q}_i^\pm$), where $i=1,\ldots, 5$, take the following form:
\begin{equation}
Z^\pm =
\left(\begin{array}{rrrrr}
Z_{11}^\pm\,\, & Z_{12}^\pm\,\, & Z_{13}^\pm\,\, & Z_{14}^\pm\,\, & Z_{15}^\pm \\
Z_{21}^\pm\,\, & Z_{22}^\pm\,\, & Z_{23}^\pm\,\, & Z_{24}^\pm\,\, & Z_{25}^\pm \\
Z_{31}^\pm\,\, & Z_{32}^\pm\,\, & Z_{33}^\pm\,\, & Z_{34}^\pm\,\, & Z_{35}^\pm \\
Z_{41}^\pm\,\, & Z_{42}^\pm\,\, & Z_{43}^\pm\,\, & Z_{44}^\pm\,\, & Z_{45}^\pm \\
Z_{51}^\pm\,\, & Z_{52}^\pm\,\, & Z_{53}^\pm\,\, & Z_{54}^\pm\,\, & Z_{55}^\pm 
\end{array}\right),
\quad
{\cal Z}^\pm
=
\left(\begin{array}{rrrrr}
 {\cal Z}_{11}^\pm  &0\,\,         &0\,\,         &0\,\,        &0\,\,  \\
 0\,\,         &{\cal Z}_{22}^\pm  &{\cal Z}_{23}^\pm  &0\,\,        &0\,\,  \\
 0\,\,         &{\cal Z}_{32}^\pm  &{\cal Z}_{33}^\pm  &0\,\,        &0\,\,  \\
 0\,\,         &0\,\,         &0\,\,         &{\cal Z}_{44}^\pm  &{\cal Z}_{45}^\pm \\
 0\,\,         &0\,\,         &0\,\,         &{\cal Z}_{54}^\pm  &{\cal Z}_{55}^\pm
\end{array}\right).
\label{MixingMatrix}
\end{equation}
For chirally invariant actions, the $Z^\pm$ matrices take the same block-diagonal structure of the $\mathcal{Z}^\pm$ matrices, as concluded in Section \ref{other_symmetries}. The corresponding renormalized  operators, $Q^{\pm,R}$ (${\cal Q}^{\pm,R}$), are defined via the equations:
\begin{equation}
Q_l^{\pm,R} = Z^\pm_{lm} \cdot Q^\pm_{m} ,\qquad
{\cal Q}^{\pm,R}_l = {\cal Z}^\pm_{lm} \cdot {\cal Q}^\pm_m,
\end{equation}
where $l,m = 1,\dots ,5$ (a sum over $m$ is implied). 

The symmetries studied in the previous section can give us specific relations between the elements of the four mixing matrices $Z^\pm$, ${\cal Z}^\pm$. As we conclude in Section \ref{other_symmetries}, operators $\mathcal{Q}_2^+$, $\mathcal{Q}_3^+$ are related to $\mathcal{Q}_2^-$, $\mathcal{Q}_3^-$ under charge conjugation; in particular, the following combinations are eigenstates of $\mathcal{C}$ with eigenvalues $C=\pm 1$: 
\begin{eqnarray}
&& (\mathcal{Q}_2^\pm)^{f_1 f_2}_{f_3 f_4} - (\mathcal{Q}_2^\mp)^{f_3 f_4}_{f_1 f_2}, \quad (\mathcal{Q}_3^\pm)^{f_1 f_2}_{f_3 f_4} + (\mathcal{Q}_3^\mp)^{f_3 f_4}_{f_1 f_2}, \qquad C= + 1, \\
&& (\mathcal{Q}_2^\pm)^{f_1 f_2}_{f_3 f_4} + (\mathcal{Q}_2^\mp)^{f_3 f_4}_{f_1 f_2}, \quad (\mathcal{Q}_3^\pm)^{f_1 f_2}_{f_3 f_4} - (\mathcal{Q}_3^\mp)^{f_3 f_4}_{f_1 f_2}, \qquad C= - 1.
\end{eqnarray}
Combinations with $C=+1$, cannot mix with the combinations with $C=-1$ and vice versa. This restriction can give us relations between elements of the plus and minus mixing matrices. As an example, we consider the $\mathcal{C}$-odd renormalized combination:
\begin{eqnarray}
    (\mathcal{Q}_2^{+,R})^{f_1 f_2}_{f_3 f_4} + (\mathcal{Q}_2^{-,R})^{f_3 f_4}_{f_1 f_2} &=& \mathcal{Z}_{22}^+ (\mathcal{Q}_2^+)^{f_1 f_2}_{f_3 f_4} + \mathcal{Z}_{23}^+ (\mathcal{Q}_3^+)^{f_1 f_2}_{f_3 f_4} + \mathcal{Z}_{22}^- (\mathcal{Q}_2^-)^{f_3 f_4}_{f_1 f_2} + \mathcal{Z}_{23}^- (\mathcal{Q}_3^-)^{f_3 f_4}_{f_1 f_2} \nonumber \\
    &=& \frac{1}{2} (\mathcal{Z}_{22}^+ + \mathcal{Z}_{22}^-) \left[(\mathcal{Q}_2^+)^{f_1 f_2}_{f_3 f_4} + (\mathcal{Q}_2^-)^{f_3 f_4}_{f_1 f_2}\right] + \frac{1}{2} (\mathcal{Z}_{22}^+ - \mathcal{Z}_{22}^-) \left[(\mathcal{Q}_2^+)^{f_1 f_2}_{f_3 f_4} - (\mathcal{Q}_2^-)^{f_3 f_4}_{f_1 f_2}\right] \nonumber \\
    &+& \frac{1}{2} (\mathcal{Z}_{23}^+ + \mathcal{Z}_{23}^-) \left[(\mathcal{Q}_3^+)^{f_1 f_2}_{f_3 f_4} + (\mathcal{Q}_3^-)^{f_3 f_4}_{f_1 f_2}\right] + \frac{1}{2} (\mathcal{Z}_{23}^+ - \mathcal{Z}_{23}^-) \left[(\mathcal{Q}_3^+)^{f_1 f_2}_{f_3 f_4} - (\mathcal{Q}_3^-)^{f_3 f_4}_{f_1 f_2}\right].
    \label{Q2_C_odd}
\end{eqnarray}
The second and third terms in the rhs of Eq. \eqref{Q2_C_odd} are $\mathcal{C}$-even and thus, must vanish, leading to: 
\begin{equation}
    \mathcal{Z}^+_{22} = \mathcal{Z}^-_{22}, \quad \mathcal{Z}^+_{23} = - \mathcal{Z}^-_{23}.
\end{equation}
Similarly, one can show that:
\begin{equation}
    \mathcal{Z}^+_{33} = \mathcal{Z}^-_{33}, \quad \mathcal{Z}^+_{32} = - \mathcal{Z}^-_{32}.
\end{equation}
These relations have been also proved in Ref.~\cite{Donini:1999sf} for a particular renormalization scheme, RI$'$, as well as in Ref.~\cite{Papinutto:2016xpq} by considering chiral symmetry. Our proof is more general since it does not refer to a specific renormalization scheme, and it is also valid in regularizations which explicitly break chiral symmetry. 

Further relations between the mixing matrices can be derived by using chiral transformations. In particular, it can be shown, by considering the eigenstates of $\mathcal{A}_2$ constructed in Table \ref{tab:chiral_eigenstates}, that the mixing matrices of the scalar operators coincide with those of the pseudoscalar operators, i.e.,
\begin{equation}
Z_{ij}^\pm = \mathcal{Z}_{ij}^\pm.
\label{ZequalZcal}
\end{equation}
This relation is only valid for chirally invariant actions.

\subsection{Gauge-invariant renormalization scheme (GIRS)}

In this work, we utilize a gauge-invariant renormalization scheme (GIRS) to determine the mixing matrices $Z^\pm$ and ${\cal Z}^\pm$. GIRS was introduced by our group in Ref.~\cite{Costa:2021iyv}, by extending previous coordinate-space prescriptions~\cite{Gimenez:2004me, Chetyrkin:2010dx, Cichy:2012is, Tomii:2018zix}, and has been applied so far in the renormalization of various composite operators in QCD (quark bilinears~\cite{Costa:2021iyv}, energy-momentum tensor~\cite{Costa:2021iyv}, $\Delta F=2$ four-quark operators~\cite{Constantinou:2024wdb}), as well as in the Supersymmetric Yang-Mills theory (gluino-glue operator~\cite{Costa:2021pfu}, Noether supercurrent~\cite{Bergner:2022see}). It is based on calculations of vacuum expectation values (vev) involving two or more gauge-invariant operators at different spacetime points, where the separations between these operators define the renormalization scale(s) of the scheme. GIRS is especially useful in lattice regularizations in which nonperturbative determinations of the mixing matrices can be obtained. Since the scheme is inherently gauge invariant, it obviates the need for gauge fixing, thereby avoiding issues related to Gribov copies in lattice numerical simulations, as well as considering mixing with gauge noninvariant operators.

When operator mixing occurs, one needs to consider a set of conditions involving more than one Green's functions of products of gauge-invariant operators. Vev must be singlets under the flavor group because any flavor rotation can be absorbed through a change of variables in the functional integral. Therefore, to obtain a nonzero vev, the tensor product of the representations of the operators in a Green's function must contain the trivial (singlet) representation. In our study, the determination of the $5 \times 5$ mixing matrices $Z^\pm$ and $\mathcal{Z}^\pm$ requires the calculation of not only two-point Green's functions with two four-quark operators, but also three-point Green's functions with one four-quark operator and two lower dimensional operators, e.g., quark bilinear operators:
\begin{equation}
 {\big({\cal O}_{\Gamma}\big)}^{f_1} {}_{f_2}\,(x) \equiv \bar{\psi}^{f_1}(x) \Gamma \psi_{f_2}(x).   
\end{equation}
There are different possible choices of these Green's functions (see below), and thus different variants of GIRS can be realized. The tensor product of two representations $R_1$, $R_2$, can contain a singlet only if $R_2$ is the conjugate of $R_1$. Thus, the two four-quark operators in the two-point Green's functions must be selected appropriately [see Eqs. (\ref{Q2pt}, \ref{calQ2pt})]. Two-point Green's functions with one four-quark operator and one quark bilinear operator are not considered since they vanish for the traceless four-quark operators studied in this work. This is understood since the tensor product of a traceless four-quark operator with a quark bilinear operator (traceless or pure trace) cannot contain a singlet representation, leading to a zero vev. On the other hand, the tensor product of one four-quark operator and two quark bilinear operators can contain at most one singlet representation. All operators in the Green's functions are placed at different spacetime points, in a way as to avoid potential contact singularities. Also, to reduce statistical noise in lattice simulations, integration (summation on the lattice) over time slices of the operator-insertion points in all Green’s functions has been employed. Although the integration does not filter out long spatial separations of the operators involved, the resulting correlators are safe from long-distance effects, if the time separations are small enough ($\ll \Lambda_{\rm QCD}$, where $\Lambda_{\rm QCD}$ is the QCD physical scale)\footnote{We are grateful to Vittorio Lubicz for identifying this issue and providing its resolution.}. This is true even for zero masses. In fact, integrating over the spatial coordinates implies that the correlators are Fourier-transformed at a fixed spatial momentum, in particular at zero momentum. These correlators then receive contributions from intermediate on-shell states with fixed momentum, and one can show that for small time separations, the dominant contribution comes from high-energy states, typically with energies on the order of the inverse of the time separations. These contributions are short-distance and thus, perturbation theory can be safely applied. The Green's functions under consideration are: 
\begin{eqnarray}
{\Big(G^{\rm 2pt}_{Q_{i}^\pm;\, Q_{j}^\pm}\Big)}^{f_1,f_2;f_1',f_2'}_{f_3,f_4;f_3'f_4'}\,(z_4) &\equiv& \int d^3 \vec{z} \ \left\langle {\big(Q_i^\pm\big)}^{f_1,f_2}_{f_3,f_4}\,(x+z)\, \left[{\big(Q_j^\pm\big)}^{f_3',f_4'}_{f_1',f_2'}\,(x)\right]^\dagger \right\rangle, \label{Q2pt} \\
{\Big(G^{\rm 3pt}_{\mathcal{O}_{\Gamma};\,Q_{i}^\pm ;\, \mathcal{O}_{\Gamma}}\Big)}^{f_1,f_2;f_1',f_2'}_{f_3,f_4;f_3'f_4'}\,(z_4,z'_4) &\equiv& \int d^3 \vec{z} \int d^3 \vec{z'} \, \left\langle {\big({\mathcal O}_{\Gamma} \big)}^{f_1'} {}_{f_3'}\,(x + z)\,\, {\big(Q_{i}^\pm\big)}^{f_1,f_2}_{f_3,f_4}\,(x)\,\, {\big({\mathcal O}_{\Gamma}\big)}^{f_2'} {}_{f_4'}\,(x - z') \right\rangle, \label{Q3pt} \\
{\Big(G^{\rm 2pt}_{\mathcal{Q}_{i}^\pm;\, \mathcal{Q}_{j}^\pm}\Big)}^{f_1,f_2;f_1',f_2'}_{f_3,f_4;f_3'f_4'}\,(z_4) &\equiv& \int d^3 \vec{z} \ \left\langle {\big(\mathcal{Q}_i^\pm\big)}^{f_1,f_2}_{f_3,f_4}\,(x+z)\, \left[{\big(\mathcal{Q}_j^\pm\big)}^{f_3',f_4'}_{f_1',f_2'}\,(x)\right]^\dagger \right\rangle, \label{calQ2pt} \\
{\Big(G^{\rm 3pt}_{\mathcal{O}_{\Gamma};\,\mathcal{Q}_{i}^\pm ;\, \mathcal{O}_{\Gamma \gamma_5}}\Big)}^{f_1,f_2;f_1',f_2'}_{f_3,f_4;f_3'f_4'}\,(z_4,z'_4) &\equiv& \int d^3 \vec{z} \int d^3 \vec{z'} \, \left\langle {\big({\mathcal O}_{\Gamma} \big)}^{f_1'} {}_{f_3'}\,(x + z)\,\, {\big(\mathcal{Q}_{i}^\pm\big)}^{f_1,f_2}_{f_3,f_4}\,(x)\,\, {\big({\mathcal O}_{\Gamma \gamma_5}\big)}^{f_2'} {}_{f_4'}\,(x - z') \right\rangle, \label{calQ3pt}
\end{eqnarray}
where $z=(\vec{z},z_4)$, $z'= (\vec{z'},z'_4)$, $z_4>0$, $z'_4>0$; $i, j$ runs from 1 to 5 and $i \leq j$; $\Gamma \in \{\openone,\, \gamma_5,\, \gamma_\mu,\, \gamma_\mu \gamma_5,\, \sigma_{\mu\nu} \}$.  The hermitian conjugate operator taken in the two-point Green's functions (Eqs. (\ref{Q2pt}, \ref{calQ2pt})) ensures that a flavor singlet representation exists and thus vev is nonzero. The representations supported by the scalar operators $Q_i^\pm$ are self-conjugate, while the representations supported by $\mathcal{Q}_i^+$ and $\mathcal{Q}_i^-$ are conjugates of each other.
The gamma matrices of the quark bilinear operators in the three-point functions have been selected in a way as to give nonzero values. Similarly, flavor indices ($f_1$, $f_2$) and ($f_1'$, $f_2'$) must be pairwise equal to ($f_3'$, $f_4'$) and ($f_3$, $f_4$), respectively. Note that all pairwise matchings of flavor indices will be equivalent (up to a group-theoretical factor), since there is only one flavor singlet that can be built out of the tensor product representations appearing in the above 4 equations. 

In order to determine a consistent and solvable set of nonperturbative renormalization conditions, we need to examine multiple choices of three-point functions with different bilinear operators; such choices define different variants of GIRS. In Ref.\cite{Constantinou:2024wdb}, we have performed a next-to-leading order (NLO) perturbative calculation of all possible Green's functions derived by Eqs. (\ref{Q2pt} -- \ref{calQ3pt}), for the operators of Case 1 (see Section \ref{sec:DF}), in dimensional regularization (DR) and we have determined NLO conversion factors from different variants of GIRS to the $\overline{\rm MS}$ scheme.

\begin{figure}[ht!]
\includegraphics[scale=0.9]{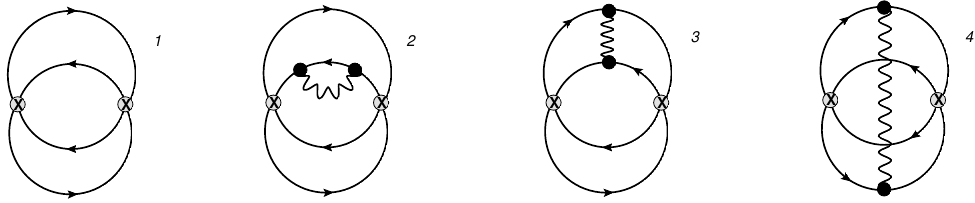} 
\caption{Feynman diagrams contributing to $\langle {\cal O}_{\Gamma \Tilde{\Gamma}} (x) \,  {\cal O}^\dagger_{\Gamma' \Tilde{\Gamma'}} (y) \rangle$, to order $\mathcal{O} (g^0)$ (diagram 1) and $\mathcal{O} (g^2)$ (the remaining diagrams). Wavy (solid) lines represent gluons (quarks). Diagrams 2 and 4 have also mirror variants.}
\label{fig4Q4Q:2pt}
\end{figure}

\begin{figure}[ht!]
\includegraphics[scale=0.8]{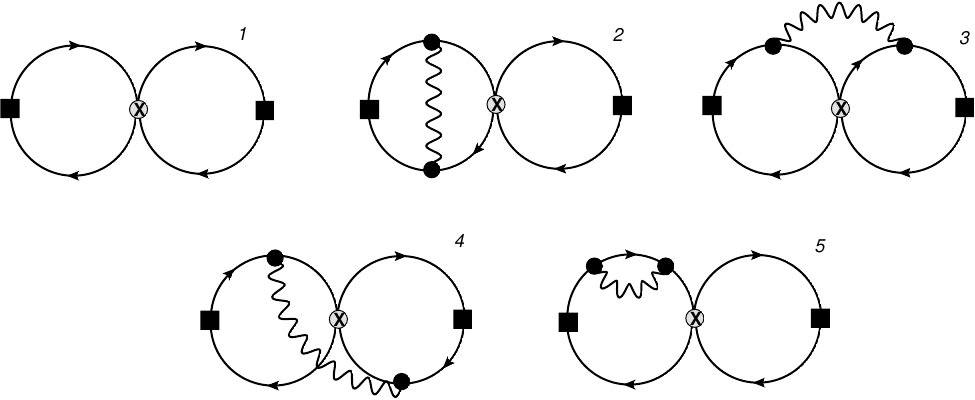} 
\caption{Feynman diagrams contributing to $\langle {\cal O}_{\Gamma'} (x) \, {\cal O}_{\Gamma \Tilde{\Gamma}} (0) {\cal O}_{\Gamma''} (y) \rangle$, to order $\mathcal{O} (g^0)$ (diagram 1) and $\mathcal{O} (g^2)$ (the remaining diagrams). Wavy (solid) lines represent gluons (quarks). A circled cross denotes the insertion of the four-quark operator, and the solid squares denote the quark bilinear operators. Diagrams 2-5 have also mirror variants. 
}
\label{fig4Q2Q2Q:3pt}
\end{figure}

The Feynman diagrams contributing to the two-point functions of the four-quark operators, to leading order (LO) ${\cal O}(g^0)$ (diagram 1) and to the NLO ${\cal O} (g^2)$ (the remaining diagrams), are shown in Fig.~\ref{fig4Q4Q:2pt}. Likewise, the Feynman diagrams contributing to the three-point Green's functions of the product of one four-quark operator and two quark bilinear operators are shown in Fig.~\ref{fig4Q2Q2Q:3pt}. For simplicity, we have not drawn separate diagrams to specify which quark/antiquark appearing in the definition of the four-quark operators is contracted in each fermion propagator. Thus, in each diagram, it is understood that all possible ways of contracting the quark/antiquark fields of the operators are summed over. The Green's functions are gauge-independent at each perturbative order; indeed, terms dependent on the gauge parameter cancel out upon summation of the Feynman diagrams.

\begin{figure}[ht!]
\includegraphics[scale=0.525]{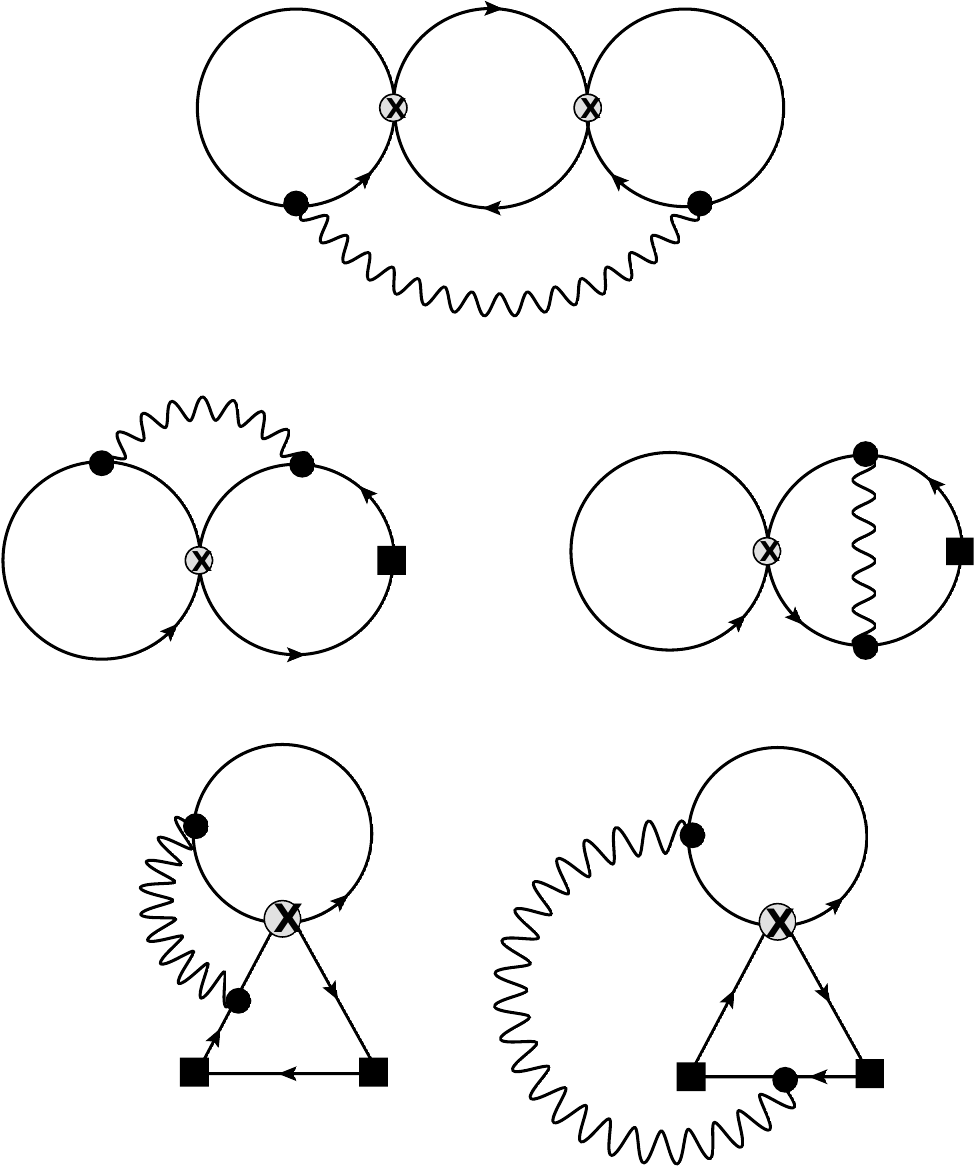} 
\caption{Representative Feynman diagrams contributing to two-point and three-point Green's functions at the next-to-leading order, $\mathcal{O}(g^2)$. Wavy (solid) lines represent gluons (quarks). The four-quark operator ${{\cal O}}_{\Gamma \Gamma'}$ is indicated by a circled cross. Bilinear operators, ${\cal O}_\Gamma$ are denoted by solid squares, while solid circles correspond to interaction vertices from the action. The first row shows examples of two-point Green's functions of the form $\langle {{\cal O}}_{\Gamma \Gamma'} (x) \,  {{\cal O}}^\dagger_{\widetilde{\Gamma} \widetilde{\Gamma}'} (y) \rangle$. The second row includes diagrams for two-point Green's functions between a four-quark operator and a bilinear, $\langle {{\cal O}}_{\Gamma \Gamma'} (x) \,  {\cal O}^\dagger_{\widetilde{\Gamma}} (y) \rangle$. The last row illustrates three-point Green's functions involving one four-quark and two bilinear operators: $\langle {\cal O}_{\widetilde{\Gamma}} (x) \, {{\cal O}}_{\Gamma \Gamma'} (y) {\cal O}_{\widetilde{\Gamma}'} (z) \rangle$. All these diagrams do not vanish in dimensional regularization (DR) or by color algebra (e.g., by Fierz rearrangements), but they do vanish when the four-quark operator is traceless due to the presence of partial traces over a traceless operator.}
\label{novanishing}
\end{figure}

For a generic flavor-group representation of the four-quark operators, i.e., operators $\widecheck{\mathcal{O}}$, $\overline{\mathcal{O}}$ and $\widehat{\mathcal{O}}$ (see Eqs.~(\ref{Op1}-\ref{Op5})), additional Feynman diagrams must be considered. In Fig.~\ref{novanishing}, we display representative Feynman diagrams contributing non-trivially to two-point and three-point Green's functions. These diagrams arise at order $\mathcal{O}(g^2)$, and illustrate cases where the contributions do not vanish in dimensional regularization or by color algebra alone (e.g., by Fierz rearrangements). However, these diagrams do vanish for the traceless representations under study ($\widecheck{\mathcal{O}}$) because they correspond to a partial trace of a traceless operator. This feature is independent of the number of gluon exchanges between quark lines. 

When performing calculations in DR, one needs to define $\gamma_5$ in $D$ dimensions. In our study, we have employed the t’Hooft-Veltman (HV) prescription~\cite{tHooft:1972tcz}, in which the following commutation/anticommutation relations of $\gamma_5$ are valid:
\begin{equation}
    \{ \gamma_5, \gamma_\mu \}=0, \ \mu=1,2,3,4, \qquad [\gamma_5, \gamma_\mu]=0, \ \mu>4.
\end{equation}
This choice does not lead to algebraic inconsistencies with $\gamma_5$-odd
fermion traces \cite{Collins:105730}, as in the naive prescription of DR (NDR). However, it violates chiral symmetry and, thus, the mixing matrices $Z^\pm$ are no longer block diagonal, and Eq. \ref{ZequalZcal} is not necessarily valid in DR with the HV prescription. However, these properties are not violated for the one-loop mixing coefficients in the $\overline{\rm MS}$ scheme with HV prescription, as concluded by our calculation. One can restore the chiral properties of the scalar operators in the HV prescription by multiplying with appropriate finite conversion factors. These factors can be extracted from our calculation to NLO. We note that in our study Lorentz indices appearing in the definition of the four-quark operators and quark bilinear operators, as well as the scales $z$, $z'$ are taken to lie in 4 instead of $D$ dimensions, which simplifies the trace algebra in this prescription. 

Another complication of DR is that four-quark operators can also potentially mix with evanescent operators, which emerge at finite values of the regulator $\varepsilon$, where $D = 4-2 \varepsilon$. In higher perturbative orders, the presence of evanescent operators must be taken into account when working with regularizations defined in $D$ dimensions \cite{Collins:105730, Dugan:1990df,Herrlich:1994kh}; in particular, one can define ``evanescent-subtracted'' operators~\cite{Lehner:2011fz,Lin:2024mws}, for a more direct matching to other regularizations. The basis of evanescent operators is arbitrary, and each choice can lead to a different renormalization scheme. For the NLO calculations in DR, all $\mathcal{O}(1/\epsilon)$ divergences can be removed from Green's functions involving the 4-fermi operators without recurrence to evanescent operators, so that the latter amount to inducing at most a further finite renormalization in $\overline{\rm MS}$. For renormalization studies of four-quark operators including evanescent operators, see Refs.~\cite{Herrlich:1994kh,Ciuchini:1997bw,Lehner:2011fz,Ciuchini:1995cd,Buhler:2023gsg,Lin:2024mws}.

\section{Renormalization conditions and Results}
\label{Results}

In this section, we construct all possible sets of renormalization conditions in GIRS based on the calculated coordinate-space Green’s functions. This results in several GIRS variants, including a democratic version discussed in Sec.~\ref{sec:democratic}. For selected variants with reduced operator mixing, we provide the conversion matrices to the $\overline{\text{MS}}$ scheme.

\subsection{GIRS renormalization conditions}
\label{conditions}

As explained in Section \ref{sec:DF}, operators that transform under the same representation of the flavor group share the same renormalization. Thus, the determination of the mixing matrices in a renormalization scheme can be implemented considering specific members of each flavor multiplet. Consequently, we construct renormalization conditions in GIRS by selecting only four-quark operators with $f_1 \notin \{f_3, f_4\} $ and $f_2 \notin \{f_3, f_4\}$ (Case 1), in which trace terms vanish. These can be $\Delta F=2$ operators ($f_1 = f_2$ or $f_3 = f_4$; see Section~\ref{sec:DF} on their possible existence) or $\Delta F=1$ operators with $(f_1, f_2, f_3, f_4)$ all different. These choices lead to the calculation of the same contractions among mass-degenerate quark/antiquark fields, but multiplied by  different prefactors; thus, they will amount to different values of the calculated Green's functions. However, they all lead to the same mixing matrix, since the factors of difference between the Green's functions of the two choices are canceled by their LO values used in the renormalization conditions (see Eqs. (\ref{cond2pt} -- \ref{cond3pconserving}, \ref{cond2ptcalQ1calQ1} -- \ref{cond3ptviol})). 
We note that the renormalization conditions can be constructed alternatively by using the operators of Case 2 or 3. The latter choices will lead to the calculation of the same Feynman diagrams (and the same contractions among mass-degenerate quark/antiquark fields); additional disconnected diagrams will be zero for the traceless $\widecheck{\mathcal{O}}$ representations. From the lattice simulation point of view,  it is not clear which option can give a better signal. However, given that Case 1 operators have no explicit trace terms, they are expected to be a preferable choice. The selected case is the one studied in our previous work~\cite{Constantinou:2024wdb}. In the following, we discuss different choices of renormalization conditions based on that work. Additionally, a ``democratic'' version is presented in Section \ref{sec:democratic}. The conditions are defined in a regularization free of evanescent operators and with explicit chiral-symmetry breaking. The flavor indices are omitted for simplicity. It is understood that they are selected in a way that the two-point and three-point Green's functions are not zero (see flavor structures in Eqs. (\ref{2ptCons} -- \ref{3ptCons}, \ref{2ptViol} -- \ref{3ptViol})).

In the case of scalar operators ($Q_i^\pm$), the mixing matrix $Z^\pm$ is $5\times5$ for both $Q_i^+$ and $Q_i^-$. Therefore, for each case, we need 25 conditions to obtain these mixing coefficients. Computing the relevant two-point Green's functions, we extract 15 conditions, and hence, we need another 10 conditions that will be extracted from the relevant three-point Green's functions. The 15 conditions in GIRS, which include two-point Green's functions, are the following: 
\begin{equation}
[G^{\rm 2pt}_{Q_i^{\pm}; Q_j^{\pm}} (t)]^{\rm GIRS} \equiv \sum_{k,l = 1}^5 (Z_{ik}^{\pm})^{\rm GIRS} (Z_{jl}^{\pm})^{\rm GIRS} \ G^{\rm 2pt}_{Q_k^{\pm}; Q_l^{\pm}} (t) = [G^{\rm 2pt}_{Q_i^{\pm};Q_j^{\pm}} (t)]^{\rm free},
\label{cond2pt}
\end{equation}
where $i, j$ run from 1 to 5 and $i \leq j$; $z_4 \equiv t$ is the GIRS renormalization scale. We have a variety of options for selecting the remaining 10 conditions involving three-point Green's functions:
\begin{equation}
[G^{\rm 3pt}_{\mathcal{O}_\Gamma;Q_i^{\pm}; \mathcal{O}_\Gamma} (t,t')]^{\rm GIRS} \equiv (Z_{\mathcal{O}_\Gamma}^{\rm GIRS})^2 \ \sum_{k=1}^5 \ (Z_{ik}^{\pm})^{\rm GIRS} \ G^{\rm 3pt}_{\mathcal{O}_\Gamma; Q_k^{\pm}; \mathcal{O}_\Gamma} (t,t') = [G^{\rm 3pt}_{\mathcal{O}_\Gamma;Q_i^{\pm}; \mathcal{O}_\Gamma} (t,t')]^{\rm free},
\label{cond3pconserving}
\end{equation}
where $i \in [1,5]$, $\Gamma \in \{\openone,\, \gamma_5,\, \gamma_m,\, \gamma_m \gamma_5,\, \sigma_{mn},\,\sigma_{m4} \}$ ($m,n$ are spatial directions), and $z_4 \equiv t$, $z'_4 \equiv t'$ are GIRS renormalization scales. A natural choice is to set $t'=t$, leading to one scale. More general relative values of $t$ and $t'$ are investigated in Ref.~\cite{Constantinou:2024wdb}. $Z_{\mathcal{O}_\Gamma}^{\rm GIRS}$ is the renormalization factor of the bilinear operator $\mathcal{O}_\Gamma$ calculated in Ref.~\cite{Costa:2021iyv}. $[G^{\rm 2pt}_{Q_i^{\pm}; Q_j^{\pm}}]^{\rm free}$ and $[G^{\rm 3pt}_{\mathcal{O}_\Gamma;Q_i^{\pm}; \mathcal{O}_\Gamma}]^{\rm free}$, appearing in the r.h.s. of Eqs. \eqref{cond2pt} and \eqref{cond3pconserving}, are the values of $G^{\rm 2pt}_{Q_i^{\pm}; Q_j^{\pm}}$ and $G^{\rm 3pt}_{\mathcal{O}_\Gamma;Q_i^{\pm}; \mathcal{O}_\Gamma}$ in the corresponding free theory, where $\mathcal{O} (g^n), \, n>0$ contributions are absent; we provide below the free-theory values, as extracted from our calculation:
\begin{eqnarray}
[G^{\rm 2pt}_{Q_i^{\pm}; Q_j^{\pm}} (t)]^{\rm free} &=& \frac{N_c}{\pi^6 |t|^9} \bigl(\delta^{f_1}{}_{f'_3} \,\delta^{f_2}{}_{f'_4} \, {\pm} \, \delta^{f_1} {}_{f'_4} \,\delta^{f_2} {}_{f'_3}\bigr) \!\bigl(\delta_{f_3} {}^{f'_1} \,\delta_{f_4} {}^{f'_2} \, {\pm} \, \delta_{f_3} {}^{f'_2} \,\delta_{f_4} {}^{f'_1}\bigr) \!\bigl(a_{ij;0}^{\pm} + a_{ij;1}^{\pm} N_c \bigr), \label{2ptCons} \\
{[G^{\rm 3pt}_{\mathcal{O}_\Gamma;Q_i^{\pm}; \mathcal{O}_{\Gamma}} (t,t)]}^{\rm free} &=& \frac{N_c}{\pi^4 t^6} \bigl(\delta^{f_1} {}_{f'_3} \,\delta^{f_2} {}_{f'_4} \, {\pm} \, \delta^{f_1} {}_{f'_4} \,\delta^{f_2} {}_{f'_3}\bigr) \!\bigl(\delta_{f_3} {}^{f'_1} \,\delta_{f_4} {}^{f'_2} \, {\pm} \, \delta_{f_3} {}^{f'_2} \,\delta_{f_4} {}^{f'_1}\bigr) \!\bigl(d_{i\Gamma;0}^{\pm} + d_{i\Gamma;1}^{\pm} N_c \bigr), \label{3ptCons}
\end{eqnarray}
where the coefficients $a_{ij;k}^{\pm}$ and $d_{i\Gamma;k}^{\pm}$ are given in Tables~\ref{tab:GFs2ptCons} and \ref{tab:GFs3ptCons}. Note that three-point functions which include temporal components of vector ($V_4$) and/or axial-vector ($A_4$) operators vanish (under integration over timeslices) and thus they are omitted from Table~\ref{tab:GFs3ptCons}, leaving only the spatial components ($\vec{V},\ \vec{A\,}$).

From Eq. \eqref{cond3pconserving}, we can obtain a total of 30 conditions with non-vanishing three-point functions containing one of the 5 scalar four-quark operators $Q_i$ and one of the 6 possible choices of quark bilinear operators $\mathcal{O}_\Gamma$ with $\Gamma \in \{\openone,\, \gamma_5,\, \gamma_m,\, \gamma_m \gamma_5,\, \sigma_{mn},\,\sigma_{m4} \}$. Thus, there are $30!/(10! \ 20!) = 30,045,015$ different choices for obtaining the remaining 10 conditions from the three-point functions. However, some choices include linearly dependent or incompatible conditions leading to infinite or no solutions, respectively. This potential incompatibility is further illustrated in Section \ref{sec:democratic}. By examining all cases in one-loop perturbation theory, we conclude that there are 205,088 choices of conditions, which give a unique solution. 

Even though all solvable systems of conditions are acceptable, it is natural to set a criterion in order to select options which have better behavior compared to others. In practice, one can choose the specific conditions that provide a more stable signal in numerical simulations. From the perturbative point of view, such a criterion can be the size of the mixing contributions. To this end, we evaluate the sum of squares of the one-loop finite terms in the off-diagonal elements of the GIRS mixing matrices (or equivalently the one-loop off-diagonal coefficients in the conversion matrices, which are given in the next subsection) for all the accepted cases, and we choose the cases with the smallest values. We found that, in general, the sums of squares among different choices are comparable. We provide below one of the options that give the minimum sum of squares, in which the following 10 renormalized three-point functions are included:
\begin{eqnarray}
&&  G^{\rm 3pt}_{S;Q_1^{\pm}; S} (t,t), \qquad G^{\rm 3pt}_{P;Q_1^{\pm}; P} (t,t), \qquad G^{\rm 3pt}_{V_i;Q_1^{\pm}; V_i} (t,t), \qquad G^{\rm 3pt}_{S;Q_2^{\pm}; S} (t,t), \qquad G^{\rm 3pt}_{P;Q_2^{\pm}; P} (t,t), \nonumber\\
&&  G^{\rm 3pt}_{S;Q_3^{\pm}; S} (t,t), \qquad G^{\rm 3pt}_{S;Q_5^{\pm}; S} (t,t), \qquad G^{\rm 3pt}_{P;Q_5^{\pm}; P} (t,t), \qquad G^{\rm 3pt}_{V_i;Q_5^{\pm}; V_i} (t,t), \qquad G^{\rm 3pt}_{A_i;Q_5^{\pm}; A_i} (t,t),
\label{selectedcondition}
\end{eqnarray}

In the case of the pseudoscalar operators ($\mathcal{Q}_i^\pm$), the $5\times5$ mixing matrix $\mathcal{Z}^\pm$ is decomposed into three blocks for both $\mathcal{Q}_i^+$ and $\mathcal{Q}_i^-$. The blocks contain: $\{\mathcal{Q}_1\}$, $\{ \mathcal{Q}_2, \mathcal{Q}_3 \}$ and $\{\mathcal{Q}_4, \mathcal{Q}_5 \}$, respectively. The first block includes only one operator, which is multiplicatively renormalizable; thus, only 1 condition is needed and can be obtained from a two-point function. The second and third blocks each contain two operators, requiring 4 conditions per block to determine the mixing coefficients. Three of them will be extracted from two-point functions, while the remaining one condition requires the calculation of a three-point Green's function. The conditions that include two-point Green's functions for each block are the following:
\begin{eqnarray}
[G^{\rm 2pt}_{\mathcal{Q}_1^{\pm}; \mathcal{Q}_1^{\pm}} (t)]^{\rm GIRS} &\equiv& [(\mathcal{Z}_{11}^{\pm})^{\rm GIRS}]^2 \ G^{\rm 2pt}_{\mathcal{Q}_1^{\pm}; \mathcal{Q}_1^{\pm}} (t) = [G^{\rm 2pt}_{\mathcal{Q}_1^{\pm}; \mathcal{Q}_1^{\pm}} (t)]^{\rm free}, \label{cond2ptcalQ1calQ1} \\
{[G^{\rm 2pt}_{\mathcal{Q}_i^{\pm}; \mathcal{Q}_j^{\pm}} (t)]}^{\rm GIRS} &\equiv& \sum_{k,l=2}^3 \ (\mathcal{Z}_{ik}^{\pm})^{\rm GIRS} (\mathcal{Z}_{jl}^{\pm})^{\rm GIRS} \ G^{\rm 2pt}_{\mathcal{Q}_k^{\pm}; \mathcal{Q}_l^{\pm}} (t) = [G^{\rm 2pt}_{\mathcal{Q}_i^{\pm}; \mathcal{Q}_j^{\pm}} (t)]^{\rm free}, \ (i,j = 2,3), \label{cond2ptviol} \qquad \\
{[G^{\rm 2pt}_{\mathcal{Q}_i^{\pm}; \mathcal{Q}_j^{\pm}} (t)]}^{\rm GIRS} &\equiv& \sum_{k,l = 4}^5 \ (\mathcal{Z}_{ik}^{\pm})^{\rm GIRS} (\mathcal{Z}_{jl}^{\pm})^{\rm GIRS} \ G^{\rm 2pt}_{\mathcal{Q}_k^{\pm}; \mathcal{Q}_l^{\pm}} (t) = [G^{\rm 2pt}_{\mathcal{Q}_i^{\pm}; \mathcal{Q}_j^{\pm}} (t)]^{\rm free}, \ (i,j = 4, 5), \qquad
\end{eqnarray}
where $i \leq j$. The possible conditions that include three-point functions for the second and third blocks are:
\begin{eqnarray}
[G^{\rm 3pt}_{\mathcal{O}_\Gamma;\mathcal{Q}_i^{\pm}; \mathcal{O}_{\Gamma \gamma_5}} (t,t')]^{\rm GIRS} &\equiv& Z_{\mathcal{O}_\Gamma}^{\rm GIRS} \ Z_{\mathcal{O}_\Gamma \gamma_5}^{\rm GIRS} \ \sum_{k=2}^3 \ (\mathcal{Z}_{ik}^{\pm})^{\rm GIRS} \ G^{\rm 3pt}_{\mathcal{O}_\Gamma; \mathcal{Q}_k^{\pm}; \mathcal{O}_{\Gamma \gamma_5}} (t,t') \nonumber \\
&=& [G^{\rm 3pt}_{\mathcal{O}_\Gamma;\mathcal{Q}_i^{\pm}; \mathcal{O}_{\Gamma \gamma_5}} (t,t')]^{\rm free}, \ (i=2 \ {\rm or} \ 3), \label{cond3ptQ2Q3}\\
{[G^{\rm 3pt}_{\mathcal{O}_\Gamma;\mathcal{Q}_i^{\pm}; \mathcal{O}_{\Gamma \gamma_5}} (t,t')]}^{\rm GIRS} &\equiv& Z_{\mathcal{O}_\Gamma}^{\rm GIRS} \ Z_{\mathcal{O}_\Gamma \gamma_5}^{\rm GIRS} \ \sum_{k=4}^5 \ (\mathcal{Z}_{ik}^{\pm})^{\rm GIRS} \ G^{\rm 3pt}_{\mathcal{O}_\Gamma; \mathcal{Q}_k^{\pm}; \mathcal{O}_{\Gamma \gamma_5}} (t,t') \nonumber \\
&=& [G^{\rm 3pt}_{\mathcal{O}_\Gamma;\mathcal{Q}_i^{\pm}; \mathcal{O}_{\Gamma \gamma_5}} (t,t')]^{\rm free}, \ (i=4 \ {\rm or} \ 5),
\label{cond3ptviol}
\end{eqnarray}
where $\Gamma \in \{\openone,\, \gamma_m,\, \sigma_{mn} \}$ ($m,\, n$ are spatial directions). As in the scalar operators, we simplify the conditions by setting $t' = t$. The free-theory values $[G^{\rm 2pt}_{\mathcal{Q}_i^{\pm}; \mathcal{Q}_j^{\pm}} ]^{\rm free}$ and $[G^{\rm 3pt}_{\mathcal{O}_\Gamma;\mathcal{Q}_i^{\pm}; \mathcal{O}_{\Gamma \gamma_5}}]^{\rm free}$, appearing in the r.h.s. of Eqs. (\eqref{cond2ptviol} -- \eqref{cond3ptviol}), are given below, as extracted from our calculation:
\begin{eqnarray}
{[G^{\rm 2pt}_{\mathcal{Q}_i^{\pm}; \mathcal{Q}_j^{\pm}} (t)]}^{\rm free} &=& \frac{N_c}{\pi^6 |t|^9} \bigl(\delta^{f_1} {}_{f'_3} \,\delta^{f_2} {}_{f'_4} \, {\pm} \, \delta^{f_1} {}_{f'_4} \,\delta^{f_2} {}_{f'_3}\bigr) \!\bigl({(-1)}^{\delta_{i2} + \delta_{i3}} \, \delta_{f_3} {}^{f'_1} \,\delta_{f_4} {}^{f'_2} \, {\pm} \, \delta_{f_3} {}^{f'_2} \,\delta_{f_4} {}^{f'_1}\bigr) \!\bigl(\tilde{a}_{ij;0}^{\pm} + \tilde{a}_{ij;1}^{\pm} N_c \bigr)\!,\label{2ptViol} \\
{[G^{\rm 3pt}_{\mathcal{O}_\Gamma;\mathcal{Q}_i^{\pm}; \mathcal{O}_{\Gamma \gamma_5}} (t,t)]}^{\rm free} &=& \frac{N_c}{\pi^4 t^6} \bigl(\delta^{f_1} {}_{f'_3} \,\delta^{f_2} {}_{f'_4} \, {\pm} \, \delta^{f_1} {}_{f'_4} \,\delta^{f_2} {}_{f'_3}\bigr) \!\bigl({(-1)}^{\delta_{i2} + \delta_{i3}} \, \delta_{f_3} {}^{f'_1} \,\delta_{f_4} {}^{f'_2} \, {\pm} \, \delta_{f_3} {}^{f'_2} \,\delta_{f_4} {}^{f'_1}\bigr) \!\bigl(\tilde{d}_{i\Gamma;0}^{\pm} + \tilde{d}_{i\Gamma;1}^{\pm} N_c \bigr)\!, \label{3ptViol}
\end{eqnarray}
where the coefficients $\tilde{a}_{ij;k}^{\pm}$ and $\tilde{d}_{i\Gamma;k}^{\pm}$ are given in Tables~\ref{tab:GFs2ptCons} and \ref{tab:GFs3ptCons}.

From Eqs. (\ref{cond3ptQ2Q3} -- \ref{cond3ptviol}), we can obtain a total of 6 alternative conditions for each of the $2\times2$ blocks $\{\mathcal{Q}_2, \mathcal{Q}_3 \}$ and $\{\mathcal{Q}_4, \mathcal{Q}_5 \}$ with non-vanishing three-point functions containing one of the 2 four-quark operators in each block, $\mathcal{Q}_i$, and one of the 3 possible choices of quark bilinear operators $\mathcal{O}_\Gamma$ with $\Gamma \in \{\openone,\, \gamma_m,\, \sigma_{mn} \}$. However, as it turns out from our calculation, only two conditions for each block give a unique solution. Applying the same criterion, as in the scalar operators, we provide below the option that gives the smallest sum of squares of the off-diagonal coefficients, in which the following renormalized three-point functions are included:
\begin{equation}
G^{\rm 3pt}_{S;\mathcal{Q}_2^{\pm}; P} (t,t), \qquad G^{\rm 3pt}_{S;\mathcal{Q}_5^{\pm}; P} (t,t). \label{selectedcondviol}
\end{equation}

\begin{table}[thb!]
  \centering
  \begin{tabular}{c|c|c|c|c|c}
  \hline
\quad $i$ \quad & \quad $j$ \quad & \quad $a_{ij;0}^{\pm}$ \quad & \quad $a_{ij;1}^{\pm}$\quad & \quad $\tilde{a}_{ij;0}^{\pm}$ \quad & \quad $\tilde{a}_{ij;1}^{\pm}$ \quad \\ [1ex] 
\hline
\hline
$1$ & $1$ & $\pm 7/32$ & $7/32$ & $\pm 7/32$ & $\phantom{-}7/32$ \\ [1ex] 
$2$ & $2$ & $0$ & $7/32$ & $0$ & $-7/32$ \\ [1ex]
$2$ & $3$ & $\mp 7/64$ & $0$ & $\pm 7/64$ & $0$ \\ [1ex]
$3$ & $3$ & $0$ & $7/128$ & $0$ & $-7/128$ \\ [1ex]
$4$ & $4$ & $\mp 7/256$ & $7/128$ & $\mp 7/256$ & $\phantom{-}7/128$ \\ [1ex]
$4$ & $5$ & $\pm 21/128$ & $0$ & $\pm 21/128$ & $0$ \\ [1ex]
$5$ & $5$ & $\pm 21/64$ & $21/32$ & $\pm 21/64$ & $21/32$ \\ [1ex]
\hline
  \end{tabular}
  \caption{Numerical values of the coefficients $a_{ij;l}^{\pm}$ and $\tilde{a}_{ij;l}^{\pm}$, appearing in Eqs.~\eqref{2ptCons}, \eqref{2ptViol}. Values of $i$ and $j$ for which all coefficients vanish have been left out.}
  \label{tab:GFs2ptCons}
\end{table}

\begin{table}[ht!]
  \centering
  \begin{tabular}{c|c|c|c|c|c}
  \hline
\quad $i$ \quad & \quad $\Gamma$ \quad & \quad $d_{i\Gamma;0}^{\pm}$ \quad & \quad $d_{i\Gamma;1}^{\pm}$ \quad & \quad $\tilde{d}_{i\Gamma;0}^{\pm}$ \quad & \quad $\tilde{d}_{i\Gamma;1}^{\pm}$ \quad \\ [1ex] 
\hline
\hline
$2$ & $S$ & $\mp 1/16$ & $0$ & $\pm 1/16$ & $0$ \\ [1ex] 
$3$ & $S$ & $0$ & $1/32$ & $0$ & $-1/32$ \\ [1ex]
$4$ & $S$ & $\mp 1/64$ & $1/32$ & $\pm 1/64$ & $-1/32$ \\ [1ex]
$5$ & $S$ & $\pm 3/32$ & $0$ & $\mp 3/32$ & $0$ \\ [1ex]
$2$ & $P$ & $\pm 1/16$ & $0$ & $0$ & $0$ \\ [1ex]
$3$ & $P$ & $0$ & $-1/32$ & $0$ & $0$ \\ [1ex]
$4$ & $P$ & $\mp 1/64$ & $1/32$ & $0$ & $0$ \\ [1ex]
$5$ & $P$ & $\pm 3/32$ & $0$ & $0$ & $0$ \\ [1ex]
$1$ & $\vec{V}$ & $\pm 1/72$ & $1/72$ & $\pm 1/72$ & $1/72$ \\ [1ex]
$2$ & $\vec{V}$ & $0$ & $1/72$ & $0$ & $-1/72$ \\ [1ex]
$3$ & $\vec{V}$ & $\mp 1/144$ & $0$ & $\pm 1/144$ & $0$ \\ [1ex]
$1$ & $\vec{A\,}$ & $\pm 1/72$ & $1/72$ & $0$ & $0$ \\ [1ex]
$2$ & $\vec{A\,}$ & $0$ & $-1/72$ & $0$ & $0$ \\ [1ex]
$3$ & $\vec{A\,}$ & $\pm 1/144$ & $0$ & $0$ & $0$ \\ [1ex]
$4$ & $T$ & $\pm 1/576$ & $0$ & $\mp 1/576$ & $0$ \\ [1ex]
$5$ & $T$ & $\pm 1/288$ & $1/144$ & $\mp 1/288$ & $-1/144$ \\ [1ex]
\hline
  \end{tabular}
  \caption{Numerical values of the coefficients $d_{i\Gamma;l}^{\pm}$ and $\tilde{d}_{i\Gamma;l}^{\pm}$, appearing in Eqs.~\eqref{3ptCons}, \eqref{3ptViol}. Values of $i$ and $\Gamma$ for which all coefficients vanish have been left out.}
  \label{tab:GFs3ptCons}
\end{table}

\subsection{Conversion matrices}

In order to arrive at the renormalized four-quark operators in the $\overline{\rm MS}$ scheme (which is the typical scheme used in phenomenological work), the conversion matrices $(C^{\pm})^{\overline{\rm MS}, \rm{GIRS}}$ and $(\Tilde{C}^{\pm})^{\overline{\rm MS}, \rm{GIRS}}$ between GIRS and $\overline{\rm MS}$ schemes are necessary:
\begin{equation}
(Z^{\pm})^{\overline{\rm MS}}  = (C^{\pm})^{\overline{\rm MS}, \rm{GIRS}} (Z^{\pm})^{\rm{GIRS}},  \qquad (\mathcal{Z}^{\pm})^{\overline{\rm MS}}  = (\Tilde{C}^{\pm})^{\overline{\rm MS}, \rm{GIRS}} (\mathcal{Z}^{\pm})^{\rm{GIRS}}.
\label{Cdef}
\end{equation}
These can be computed only perturbatively due to the very nature of $\overline{\rm MS}$. Being regularization-independent, they are evaluated more easily in DR. The conversion matrices, along with the lattice mixing matrices in GIRS, calculated nonperturbatively, allow the extraction of the lattice mixing matrices in the $\overline{\rm MS}$ scheme. 

We calculate the one-loop contributions to the conversion matrices by implementing all possible sets of conditions in GIRS. These can be extracted by rewriting the GIRS conditions, given in the previous subsection, in terms of the conversion matrices:
\bea
{[G^{\rm 2pt}_{Q_i^{\pm}; Q_j^{\pm}} (t)]}^\MSbar &=& \sum_{k,l = 1}^5 (C_{ik}^{\pm})^{\MSbar, {\rm GIRS}} \ (C_{jl}^{\pm})^{\MSbar, {\rm GIRS}} \ [G^{\rm 2pt}_{Q_k^{\pm };Q_l^{\pm}} (t)]^{\rm free}, \\
{[G^{\rm 3pt}_{\mathcal{O}_\Gamma; Q_i^{\pm}; \mathcal{O}_\Gamma} (t,t)]}^\MSbar &=& (C_{\mathcal{O}_\Gamma}^{\MSbar,{\rm GIRS}})^2 \ \sum_{k=1}^5 \ (C_{ik}^{\pm})^{\MSbar, {\rm GIRS}} \ [G^{\rm 3pt}_{\mathcal{O}_\Gamma;Q_k^{\pm}; \mathcal{O}_\Gamma} (t,t)]^{\rm free}, \\
{[G^{\rm 2pt}_{\mathcal{Q}_i^{\pm}; \mathcal{Q}_j^{\pm}} (t)]}^\MSbar &=& \sum_{k,l = 1}^5 (\Tilde{C}_{ik}^{\pm})^{\MSbar, {\rm GIRS}} \ (\Tilde{C}_{jl}^{\pm})^{\MSbar, {\rm GIRS}} \ [G^{\rm 2pt}_{\mathcal{Q}_k^{\pm};\mathcal{Q}_l^{\pm}} (t)]^{\rm free}, \\
{[G^{\rm 3pt}_{\mathcal{O}_\Gamma; \mathcal{Q}_i^{\pm}; \mathcal{O}_{\Gamma \gamma_5}} (t,t)]}^\MSbar  &=& (C_{\mathcal{O}_\Gamma}^{\MSbar,{\rm GIRS}}) \ (C_{\mathcal{O}_{\Gamma \gamma_5}}^{\MSbar,{\rm GIRS}}) \ \sum_{k=1}^5 \ (\Tilde{C}_{ik}^{\pm})^{\MSbar, {\rm GIRS}} \  [G^{\rm 3pt}_{\mathcal{O}_\Gamma;\mathcal{Q}_k^{\pm}; \mathcal{O}_{\Gamma \gamma_5}} (t,t)]^{\rm free},
\eea
where $C_{\mathcal{O}_\Gamma}^{\MSbar,{\rm GIRS}}$ is the conversion factor of the quark bilinear operator $\mathcal{O}_\Gamma$ calculated by us to one loop in Ref.~\cite{Costa:2021iyv}:
\bea
C_S^{\MSbar,{\rm GIRS}} &=& 1 + \frac{g_\MSbar^2 \ C_F}{16 \pi^2} \ \left(-\frac{1}{2} + 3 \ln (\bar{\mu}^2 t^2) + 6 \gamma_E\right) + \mathcal{O} (g_\MSbar^4), \\
C_P^{\MSbar,{\rm GIRS}} &=& 1 + \frac{g_\MSbar^2 \ C_F}{16 \pi^2} \ \left(\frac{15}{2} + 3 \ln (\bar{\mu}^2 t^2) + 6 \gamma_E\right) + \mathcal{O} (g_\MSbar^4), \\
C_V^{\MSbar,{\rm GIRS}} &=& 1 + \frac{g_\MSbar^2 \ C_F}{16 \pi^2} \ \frac{3}{2} + \mathcal{O} (g_\MSbar^4), \\
C_A^{\MSbar,{\rm GIRS}} &=& 1 + \frac{g_\MSbar^2 \ C_F}{16 \pi^2} \ \frac{11}{2} + \mathcal{O} (g_\MSbar^4), \\
C_T^{\MSbar,{\rm GIRS}} &=& 1 + \frac{g_\MSbar^2 \ C_F}{16 \pi^2} \ \left(\frac{25}{6} - \ln (\bar{\mu}^2 t^2) - 2 \gamma_E\right) + \mathcal{O} (g_\MSbar^4).
\eea
($\gamma_E$ stands for Euler's constant.) Note that the conversion matrix $(\Tilde{C}^{S \pm 1})^{\MSbar, {\rm GIRS}}$ has the block diagonal form of $\mathcal{Z}^{\pm}$ (see Eq.~\eqref{MixingMatrix}). We provide below our results for the selected options presented in Eqs. \eqref{selectedcondition} and \eqref{selectedcondviol}:
\begin{eqnarray}
    (C_{ij}^{\pm})^{\MSbar, {\rm GIRS}} &=& \delta_{ij} + \frac{g_\MSbar^2}{16 \pi^2} \sum_{k=-1}^{+1} \left[ g_{ij;k}^\pm + \left(\ln (\bar{\mu}^2 t^2) + 2 \gamma_E \right) \ h_{ij;k}^\pm \right] N_c^k + \mathcal{O} (g_\MSbar^4),
    \label{Ccons} \\
    (\Tilde{C}_{ij}^{\pm})^{\MSbar, {\rm GIRS}} &=& \delta_{ij} + \frac{g_\MSbar^2}{16 \pi^2} \sum_{k=-1}^{+1} \left[ \tilde{g}_{ij;k}^\pm + \left(\ln (\bar{\mu}^2 t^2) + 2 \gamma_E \right) \ \tilde{h}_{ij;k}^\pm \right] N_c^k + \mathcal{O} (g_\MSbar^4),
    \label{Cviol}
\end{eqnarray}
where the coefficients $g_{ij;k}^{\pm}$, $h_{ij;k}^{\pm}$, $\tilde{g}_{ij;k}^{\pm}$ and $\tilde{h}_{ij;k}^{\pm}$ are given in Tables \ref{tab:CmatrixCons} and \ref{tab:CmatrixViol}. For other GIRS variants, we refer the reader to the one-loop results presented in our previous work~\cite{Constantinou:2024wdb}. 

A by-product of our calculation is the anomalous dimension matrices in GIRS at NLO for the four-quark operators under consideration. The full results and further details can be found in Ref.~\cite{Constantinou:2024wdb}.

\begin{table}[ht!]
  \centering
  \begin{tabular}{c|c|c|c|c|c|c|c}
  \hline
\quad $i$ \quad & \quad $j$ \quad & \quad $g_{ij;-1}^{\pm}$ \quad & \quad $g_{ij;0}^{\pm}$ \quad & \quad $g_{ij;+1}^{\pm}$ \quad & \quad $h_{ij;-1}^{\pm}$ \quad & \quad $h_{ij;0}^{\pm}$ \quad & \quad $h_{ij;+1}^{\pm}$ \\ [1ex] 
\hline
\hline
$1$ & $1$ & $-869/140$ & $\pm 379/140$ & $7/2$ & $3$ & $\mp 3$ & $0$ \\
$1$ & $2$ & $2$ & $\mp (723/280 - 6 \ln (2))$ & $-2$ & $0$ & $0$ & $0$ \\
$1$ & $3$ & $-723/140 + 12 \ln (2)$ & $0$ & $0$ & $0$ & $0$ & $0$ \\
$1$ & $4$ & $-4$ & $\pm 4$ & $0$ & $0$ & $0$ & $0$ \\
$1$ & $5$ & $-2$ & $\pm 2$ & $0$ & $0$ & $0$ & $0$ \\
$2$ & $1$ & $397/280 + 6 \ln (2)$ & $\pm (163/280 -6 \ln (2))$ & $-2$ & $0$ & $0$ & $0$ \\
$2$ & $2$ & $-9/2$ & $\pm 2$ & $7/2$ & $-3$ & $0$ & $0$ \\
$2$ & $3$ & $4$ & $\mp 2$ & $0$ & $0$ & $\mp 6$ & $0$ \\
$2$ & $4$ & $4$ & $\pm 8$ & $0$ & $0$ & $0$ & $0$ \\
$2$ & $5$ & $-2$ & $0$ & $0$ & $0$ & $0$ & $0$ \\
$3$ & $1$ & $-1$ & $\pm 1$ & $0$ & $0$ & $0$ & $0$ \\
$3$ & $2$ & $1$ & $\pm 99/280$ & $0$ & $0$ & $0$ & $0$ \\
$3$ & $3$ & $-38/35$ & $\pm 2$ & $251/140$ & $-3$ & $0$ & $3$ \\
$3$ & $4$ & $4$ & $\pm 239/280$ & $-321/140$ & $0$ & $0$ & $0$ \\
$3$ & $5$ & $0$ & $\mp 239/560$ & $0$ & $0$ & $0$ & $0$ \\
$4$ & $1$ & $-1$ & $\pm 1$ & $0$ & $0$ & $0$ & $0$ \\
$4$ & $2$ & $1$ & $\mp 239/280$ & $0$ & $0$ & $0$ & $0$ \\
$4$ & $3$ & $4$ & $\pm 2$ & $-799/140$ & $0$ & $0$ & $0$ \\
$4$ & $4$ & $-307/112 + 3 \ln (2)$ & $\pm 169/140$ & $251/140$ & $-3$ & $\mp 3$ & $3$ \\
$4$ & $5$ & $-269/480 + 1/2 \ \ln (2)$ & $\pm (869/1680 -\ln (2))$ & $0$ & $1$ & $\mp 1/2$ & $0$ \\
$5$ & $1$ & $-6$ & $\pm 6$ & $0$ & $0$ & $0$ & $0$ \\
$5$ & $2$ & $-6$ & $0$ & $0$ & $0$ & $0$ & $0$ \\
$5$ & $3$ & $0$ & $\mp 12$ & $0$ & $0$ & $0$ & $0$ \\
$5$ & $4$ & $-269/40 + 6 \ln (2)$ & $\mp (29/140 - 12 \ln (2))$ & $0$ & $12$ & $\pm 6$ & $0$ \\
$5$ & $5$ & $-1229/240-3 \ln (2)$ & $\pm 309/140$ & $1709/420$ & $1$ & $\mp 3$ & $-1$ \\
      \hline
  \end{tabular}
  \caption{Numerical values of the coefficients $g_{ij;k}^{\pm}$, $h_{ij;k}^{\pm}$ appearing in Eq.~ \eqref{Ccons}.}
  \label{tab:CmatrixCons}
\end{table}

\begin{table}[ht!]
  \centering
  \begin{tabular}{c|c|c|c|c|c|c|c}
  \hline
\quad $i$ \quad & \quad $j$ \quad & \quad $\tilde{g}_{ij;-1}^{\pm}$ \quad & \quad $\tilde{g}_{ij;0}^{\pm}$ \quad & \quad $\tilde{g}_{ij;+1}^{\pm}$ \quad & \quad $\tilde{h}_{ij;-1}^{\pm}$ \quad & \quad $\tilde{h}_{ij;0}^{\pm}$ \quad & \quad $\tilde{h}_{ij;+1}^{\pm}$ \\ [1ex] 
\hline
\hline
$1$ & $1$ & $-869/140$ & $\pm 379/140$ & $7/2$ & $3$ & $\mp 3$ & $0$ \\
$2$ & $2$ & $-9/2$ & $0$ & $7/2$ & $-3$ & $0$ & $0$ \\
$2$ & $3$ & $0$ & $\mp 2$ & $0$ & $0$ & $\mp 6$ & $0$ \\
$3$ & $2$ & $0$ & $\pm 99/280$ & $0$ & $0$ & $0$ & $0$ \\
$3$ & $3$ & $-38/35$ & $0$ & $251/140$ & $-3$ & $0$ & $3$ \\
$4$ & $4$ & $-307/112 + 3 \ln (2)$ & $\pm 169/140$ & $251/140$ & $-3$ & $\mp 3$ & $3$ \\
$4$ & $5$ & $-269/480 + 1/2 \ \ln (2)$ & $\pm (869/1680 -\ln (2))$ & $0$ & $1$ & $\mp 1/2$ & $0$ \\
$5$ & $4$ & $-269/40 + 6 \ln (2)$ & $\mp (29/140 - 12 \ln (2))$ & $0$ & $12$ & $\pm 6$ & $0$ \\
$5$ & $5$ & $-1229/240-3 \ln (2)$ & $\pm 309/140$ & $1709/420$ & $1$ & $\mp 3$ & $-1$ \\
\hline
  \end{tabular}
  \caption{Numerical values of the coefficients $\tilde{g}_{ij;k}^{\pm}$, $\tilde{h}_{ij;k}^{\pm}$ appearing in Eq.~\eqref{Cviol}.}
  \label{tab:CmatrixViol}
\end{table}

\subsection{A ``democratic'' implementation of GIRS for operator mixing}
\label{sec:democratic}

As discussed above, in the presence of operator mixing there is a vast set of choices for alternative renormalization conditions which can be employed in order to establish the mixing matrix. In this subsection we propose a particular, natural choice which treats all mixing operators on an equal footing. Our approach can be generally applied to any case of mixing; as a by-product, it also has the advantage of providing an explicit solution of the quadratic coupled equations stemming from the renormalization conditions, thus leading directly to the values of the elements of the mixing matrix.

As a prototype case involving $n$ operators, we will consider $Q_i^+ \ (i=1,\ldots,5)$. A total of $n^2$ conditions are required to determine the elements of the $n{\times}n$ mixing matrix $Z$\,; the most natural choice for $n(n+1)/2$ of these conditions entails the two-point Green's functions defined in Eq.~(\ref{Q2pt}) and denoted here as $G_{ij}$ for brevity. In concise matrix notation, Eq.~(\ref{cond2pt}) reads:
\be
Z \, G \, Z^T = G_{\rm{free}}
\ee
One can easily check that, by its definition, $G$ is a real and symmetric matrix\footnote{To this end, one must also make use of the fact that the vacuum expectation value in $G$, being necessarily a flavor singlet, forces the 8 implicit flavor indices of $G$ to be pairwise equal.} and it is positive semidefinite; these properties hold both nonperturbatively and to any order in perturbation theory. Therefore, the square root of $G$, with nonnegative eigenvalues, is uniquely defined, and it is a real symmetric matrix. Defining:
\be
Z' \equiv G_{\rm{free}}^{-1/2} \, Z \, G^{1/2},\qquad \text{there follows: }\ Z' \, Z'^T = \openone\,.
\label{Z'}
\ee

Thus the conditions stemming from two-point Green's functions simply dictate that $Z'$ be an orthogonal matrix: $Z'\in SO(n)$. This requirement indeed corresponds to $n(n+1)/2$ constraints on the elements of $Z'$, leaving another $n(n-1)/2$ degrees of freedom (same as the dimension of $SO(n)$\,) to be determined. For this determination, we turn to the three-point functions, defined in Eq.~(\ref{Q3pt}). For brevity, we will denote them as $G^{(3)}_{i,\alpha}$, where $i = 1,\ldots,n$ runs over the number of mixing operators ($n=5$ in the case at hand) and $\alpha = 1,\ldots,A$ ($A\geq n$) runs over the different types of independent three-point Green's functions under consideration; in our case, $A=6$, corresponding to the bilinears: $\Gamma \in \{\openone,\, \gamma_5,\, \gamma_m,\, \gamma_m \gamma_5,\, \sigma_{mn},\, \sigma_{m4} \}$ ($m,\, n$ are spatial directions). In terms of $G^{(3)}_{i,\alpha}$, and of the bilinear renormalization functions $Z_\alpha$\,, Eq.~(\ref{cond3pconserving}) reads:
\begin{equation}
Z_\alpha^2 \ \sum_{k=1}^n \ Z_{ik}\, G^{(3)}_{k,\alpha}  = G^{(3)\,\text{free}}_{i,\alpha}.
\label{cond3pgeneric}\end{equation}
We reiterate that, out of the $n\times A$ conditions embodied in this equation, only a choice of $n(n-1)/2$ mutually compatible conditions must be imposed, in order to arrive at a unique solution for the elements of the matrix $Z$. Defining further the rectangular matrix:
\begin{equation}
\tilde G^{(3)}_{i,\alpha} \equiv Z_\alpha^2 \ \sum_{k=1}^n \ \bigl(G^{-1/2}\bigr)_{ik}\, G^{(3)}_{k,\alpha} \, ,
\label{G3tilde}\end{equation}
Eq.~(\ref{cond3pgeneric}) simply becomes:
\begin{equation}
Z'\, \tilde G^{(3)}  = \tilde G^{(3)\,\text{free}}.
\end{equation}
Given that $Z'$ is an orthogonal matrix, it leaves invariant the lengths of each of the $A$ columns in $\tilde G^{(3)}$, as well as the inner product between any two columns. It is thus clear that these lengths and inner products cannot be compelled to become equal to their LO (``free'') values, through renormalization. This explains why, among the $30!/(10! \ 20!)$ sets of choices of renormalization conditions previously mentioned [more generally: $\binom{n\times A}{n(n-1)/2}$ ] only a small percentage corresponds to mutually compatible, non-overdetermined conditions. 

Consequently, denoting by $\tilde G^{(3)}_\alpha,\ \alpha = 1,\ldots,A$\,, each of the n-component columns of $\tilde G^{(3)}$, one could require that at most $n-1$ of the components of one such column become equal to their LO value upon renormalization. [See, e.g., conditions~(\ref{selectedcondition}), involving three-point Green's functions built out of the scalar bilinear (components of the first column of $G^{(3)}$) and only 4 out 5 $Q_i$ operators ($Q_1,\,Q_2,\,Q_3,\,Q_5$).] 

\medskip
Alternatively, an implementation of GIRS which is ``democratic" with respect to the mixing operators is as follows: 
\begin{itemize}
    \item Require that one column of $\tilde G^{(3)}$ (say the first one, $\tilde G^{(3)}_1$, which involves scalar bilinears) become, upon renormalization, parallel to its LO counterpart. This fixes $n-1$ out of the $n(n-1)/2$ parameters of $Z'$.
    \item For another column of $\tilde G^{(3)}$ (say $\tilde G^{(3)}_2$, which involves pseudoscalar bilinears), require that its components perpendicular to the first column become, upon renormalization, parallel to their LO counterpart. This fixes $n-2$ further parameters of $Z'$.
    \item Continue likewise for a $3^\text{rd},\ldots,(n{-}1)^\text{th}$ column, requiring that their components perpendicular to all previous colums become, upon renormalization, parallel to their LO counterparts. This fixes all parameters of $Z'$, and is compatible with $Z'$ belonging to $O(n)$.
\end{itemize}
In order to carry out this implementation, and deduce $Z'$, we define recursively an orthonormal basis of $n$-component vectors stemming from the above columns, using a Gram-Schmidt orthogonalization procedure: 
\be
\widehat e_1 \equiv \tilde G^{(3)}_1 \,/\,\mathcal{N}_1\,,\quad \widehat e_2 \equiv \Bigl(\tilde G^{(3)}_2 - (\tilde G^{(3)}_2 \cdot \widehat e_1) \,\widehat e_1 \Bigr) \,/\,\mathcal{N}_2\,, \quad \ldots\quad
\widehat e_i \equiv \Bigl(\tilde G^{(3)}_i - \sum_{j=1}^{i-1}(\tilde G^{(3)}_i \cdot \widehat e_j) \,\widehat e_j \Bigr) \,/\,\mathcal{N}_i \qquad (i=1,\ldots,n),
\ee
where the normalization factors $\mathcal{N}_i$ are such that\footnote{Should any normalization factor vanish, it would signal that the corresponding column of $G^{(3)}$ does not constitute an independent set of Green's functions and should thus be discarded.}: $|\widehat{e}_i| = 1$. In the same fashion, starting from $\tilde G^{(3)\,\text{free}}$ we define the LO counterparts $\widehat e_i^\text{ free}$. Note that the elements of the last vector $\widehat{e}_n$ can be exclusively expressed (up to a sign) in terms of the elements of $\widehat{e}_j$, $(j<n)$ without the need to compute the $\tilde{G}_n^{(3)}$; the sign can be determined from the corresponding LO counterpart $\widehat{e}_n^{\rm\ free}$:
\begin{equation}
    \sum_{j=1}^n {[(\widehat{e}_j)_i]}^2 = 1 \quad\Rightarrow\quad (\widehat{e}_n)_i = {\rm sgn} [(\widehat{e}^{\rm \ free}_n)_i] \ \left[{1 - \sum_{j=1}^{n-1} {[(\widehat{e}_j)_i]}^2}\right]^{1/2}.
\end{equation}
The mixing matrix $Z'$ then simply reads:
\be
Z' = \sum_{i=1}^n \widehat e_i^\text{ free} \otimes \widehat e_i^{\ T}.
\ee
It is straightforward to check that $Z'$ is indeed orthogonal and that it induces the democratic renormalization prescription described above. Combining with Eq.~(\ref{Z'}), the final result for the mixing matrix $Z$ is:
\be
Z \equiv G_{\rm{free}}^{1/2} \, Z' \, G^{-1/2}.
\ee

Applying the criterion described in the previous section, we present below one of the cases that give the smallest mixing (or, more generally, NLO) contributions:
\begin{equation}
    \Big((\tilde{G}_1^{(3),\pm})_i, (\tilde{G}_2^{(3),\pm})_i, (\tilde{G}_3^{(3),\pm})_i, (\tilde{G}_4^{(3),\pm})_i, (\tilde{G}_5^{(3),\pm})_i \Big)  = \Big(\tilde{G}^{\rm 3pt}_{S;Q_i^\pm;S}, \tilde{G}^{\rm 3pt}_{P;Q_i^\pm;P}, \tilde{G}^{\rm 3pt}_{V_m;Q_i^\pm;V_m}, \tilde{G}^{\rm 3pt}_{A_m;Q_i^+;A_m},  \tilde{G}^{\rm 3pt}_{T_{mn};Q_i^\pm;T_{mn}}\Big).
    \label{selectcondition_demGIRS}
\end{equation}
It does not really matter which tensor bilinear ($T_{mn}$ or $T_{m4}$) is being used as the fifth set of the three-point functions, since the computation of the last set is not necessary, as explained above. In this sense, this choice can be preferable given that three-point functions with tensor bilinear operators are expected to be more noisy in simulations. We note that using three-point Green's functions with both tensor bilinears $T_{mn}$ and $T_{m4}$ does not lead to a compatible set of conditions. Comparing with the conventional implementation of GIRS, this version gives a conversion matrix closer to the unit matrix and, thus, its perturbative evaluation is more trustworthy. In the democratic version (``D-GIRS'') our results for this conversion matrix at NLO are given below, for $N_c=3$: 
\begin{equation}
    (C_{ij}^{\pm})^{\MSbar, {\rm D-GIRS}} = \delta_{ij} + \frac{g_\MSbar^2}{16 \pi^2} \left[ g_{ij}^{D,\pm} + h_{ij}^{D,\pm} \ \ln (\bar{\mu}^2 t^2) \right] + \mathcal{O} (g_\MSbar^4),
    \label{Cconsdem}
\end{equation}
where $g_{ij}^{D,\pm}$ and $h_{ij}^{D,\pm}$ are given in Table \ref{tab:CmatrixConsdem}.

Whenever mixing involves only two operators, as is the case with $\{ \mathcal{Q}_2, \mathcal{Q}_3 \}$ and with $\{\mathcal{Q}_4, \mathcal{Q}_5 \}$, this procedure simplifies considerably. $Z'$, belonging to $O(2)$, can be written as:
\be
Z' = 
\begin{pmatrix*}[r]
\cos\theta & - \sin\theta\\
\sin\theta & \cos\theta
\end{pmatrix*}.
\label{Z'2x2}
\ee
In accordance with the above implementation, requiring  $Z'\,\tilde G^{(3)}_1$ to be parallel to $\tilde G^{(3)\,\text{free}}_1$ leads to\footnote{Due to the fact that, in this case, the two bilinears appearing in the 3-point Green's function differ by $\gamma_5$, the quantity $Z_\alpha^2$ in Eq.~(\ref{G3tilde}) must be replaced by the product of renormalization functions for each of the two bilinears.}:
\be
\bigl(\tilde G^{(3)\,\text{free}}_1\bigr)^T Z'\, \tilde G^{(3)}_1
= \bigl|\tilde G^{(3)\,\text{free}}_1 \bigr|\,\bigl|\tilde G^{(3)}_1\bigr| \quad \Rightarrow \quad \cos\theta = \frac{\tilde G^{(3)\,\text{free}}_1 \cdot \tilde G^{(3)}_1}{\bigl|\tilde G^{(3)\,\text{free}}_1 \bigr|\,\bigl|\tilde G^{(3)}_1\bigr|}\,, \quad \sin\theta = \frac{\tilde G^{(3)\,\text{free}}_{21} \, \tilde G^{(3)}_{11}-\tilde G^{(3)\,\text{free}}_{11} \, \tilde G^{(3)}_{21} }{\bigl|\tilde G^{(3)\,\text{free}}_1 \bigr|\,\bigl|\tilde G^{(3)}_1\bigr|}\,.
\ee

In this case, using our criterion, a possible choice of Green's function is: 
\begin{equation}
(\tilde{G}_1^{(3),\pm})_i = \tilde{G}^{\rm 3pt}_{S;\mathcal{Q}_i^\pm;P} 
\label{selectcondition_demGIRS_viol}
\end{equation}
for both $2\times 2$ sets. Our results for the conversion matrix of the selected case at NLO are given below, for $N_c=3$: 
\begin{equation}
    (\tilde{C}_{ij}^{\pm})^{\MSbar, {\rm D-GIRS}} = \delta_{ij} + \frac{g_\MSbar^2}{16 \pi^2} \left[ \tilde{g}_{ij}^{D,\pm} + \tilde{h}_{ij}^{D,\pm} \ \ln (\bar{\mu}^2 t^2) \right] + \mathcal{O} (g_\MSbar^4),
    \label{Cvioldem}
\end{equation}
where $\tilde{g}_{ij}^{D,\pm}$ and $\tilde{h}_{ij}^{D,\pm}$ are given in Table \ref{tab:CmatrixVioldem}.

\begin{table}[ht!]
  \centering
  \begin{tabular}{r|r|r|r|r|r}
  \hline
\quad $i$ \quad & \quad $j$ \quad & \quad $g_{ij}^{D,+}$ \quad & \quad $h_{ij}^{D,+}$ \quad & \quad $g_{ij}^{D,-}$ \quad & \quad $h_{ij}^{D,-}$ \\ [1ex] 
\hline
\hline
$1$ & $1$ & $8.829233$ & $-2$ & $10.341535$ & $4$ \\
$1$ & $2$ & $-6.617985$ & $0$ & $-5.888148$ & $0$ \\
$1$ & $3$ & $-0.856435$ & $0$ & $0.369876$ & $0$ \\
$1$ & $4$ & $8/3$ & $0$ & $-16/3$ & $0$ \\
$1$ & $5$ & $4/3$ & $0$ & $-8/3$ & $0$ \\
$2$ & $1$ & $-5.810232$ & $0$ & $-1.926914$ & $0$ \\
$2$ & $2$ & $10.414616$ & $-1$ & $6.414616$ & $-1$ \\
$2$ & $3$ & $-4.178969$ & $-6$ & $6.845636$ & $6$ \\
$2$ & $4$ & $9.993402$ & $0$ & $-6.267724$ & $0$ \\
$2$ & $5$ & $-2/3$ & $0$ & $-2/3$ & $0$ \\
$3$ & $1$ & $2/3$ & $0$ & $-4/3$ & $0$ \\
$3$ & $2$ & $-1/6$ & $0$ & $5/6$ & $0$ \\
$3$ & $3$ & $15.683070$ & $8$ & $11.683070$ & $8$ \\
$3$ & $4$ & $-11.656770$ & $0$ & $-10.068253$ & $0$ \\
$3$ & $5$ & $0$ & $0$ & $0$ & $0$ \\
$4$ & $1$ & $2/3$ & $0$ & $-4/3$ & $0$ \\
$4$ & $2$ & $1/3$ & $0$ & $1/3$ & $0$ \\
$4$ & $3$ & $-7.841580$ & $0$ & $-13.364817$ & $0$ \\
$4$ & $4$ & $11.540368$ & $5$ & $16.344397$ & $11$ \\
$4$ & $5$ & $0.057895$ & $-1/6$ & $0.710525$ & $5/6$ \\
$5$ & $1$ & $4$ & $0$ & $-8$ & $0$ \\
$5$ & $2$ & $-2$ & $0$ & $-2$ & $0$ \\
$5$ & $3$ & $-11.009897$ & $0$ & $12.598414$ & $0$ \\
$5$ & $4$ & $10.442128$ & $10$ & $-8.222530$ & $-2$ \\
$5$ & $5$ & $6.069376$ & $-17/3$ & $8.289952$ & $1/3$ \\
      \hline
  \end{tabular}
  \caption{Numerical values of the coefficients $g_{ij}^{D,\pm}$, $h_{ij}^{D,\pm}$ appearing in Eq.~\eqref{Cconsdem}.}
  \label{tab:CmatrixConsdem}
\end{table}

\begin{table}[ht!]
  \centering
  \begin{tabular}{r|r|r|r|r|r}
  \hline
\quad $i$ \quad & \quad $j$ \quad & \quad $\tilde{g}_{ij}^{D,+}$ \quad & \quad $\tilde{h}_{ij}^{D,+}$ \quad & \quad $\tilde{g}_{ij}^{D,-}$ \quad & \quad $\tilde{h}_{ij}^{D,-}$ \\ [1ex] 
\hline
\hline
$1$ & $1$ & $8.829233$ & $-2$ & $10.341535$ & $4$ \\
$2$ & $2$ & $8.414616$ & $-1$ & $8.414616$ & $-1$ \\
$2$ & $3$ & $-5.512302$ & $-6$ & $5.512302$ & $6$ \\
$3$ & $2$ & $-1/2$ & $0$ & $1/2$ & $0$ \\
$3$ & $3$ & $13.683070$ & $8$ & $13.683070$ & $8$ \\
$4$ & $4$ & $11.540368$ & $5$ & $16.344397$ & $11$ \\
$4$ & $5$ & $0.057895$ & $-1/6$ & $0.710525$ & $5/6$ \\
$5$ & $4$ & $10.442128$ & $10$ & $-8.222530$ & $-2$ \\
$5$ & $5$ & $6.069376$ & $-17/3$ & $8.289952$ & $1/3$ \\
\hline
  \end{tabular}
  \caption{Numerical values of the coefficients $\tilde{g}_{ij}^{D,\pm}$, $\tilde{h}_{ij}^{D,\pm}$ appearing in Eq.~\eqref{Cvioldem}.}
  \label{tab:CmatrixVioldem}
\end{table}

It remains to be investigated whether this ``democratic'' variant of GIRS can lead to a more controlled numerical error originating from the nonperturbative evaluation of $\tilde G^{(3)}_{i,\alpha}$\,. 

\section{Summary -- Future Plans}
\label{sec:conclusions}
In this work, we present a systematic classification and formulation for the renormalization of a complete set of scalar and pseudoscalar four-quark operators that do not mix with lower-dimensional operators,  due to their transformation properties under the flavor symmetry group. These operators play an important role in phenomenological studies of flavor-changing processes. Employing the Gauge-Invariant Renormalization Scheme (GIRS), we computed two-point and three-point Green’s functions involving four-quark operators and quark bilinears in coordinate space. GIRS is advantageous for lattice regularizations because its gauge-invariant nature allows the nonperturbative determination of the mixing matrices without gauge fixing, avoiding Gribov-copy issues and gauge-noninvariant operator mixing. The operator classification was carried out for a general number of flavors $N_f$, and our analysis extended beyond operators with vanishing trace parts to include a broad class of $\Delta F = 1$ and $\Delta F = 0$ operators belonging to the same flavor representations. 

To address operator mixing, we formulate suitable renormalization conditions that can be applied in both continuum and lattice regularizations. In principle, many such conditions can be defined by choosing different subsets of three-point functions (since the total number of conditions does not require using all available three-point functions) leading to different variants of GIRS. In our study, we systematically examine all possible variants within continuum perturbation theory and identify the choices that yield a definite, self-consistent solution. A selected version, which has small off-diagonal contributions to the mixing matrix is provided in Eqs. (\ref{selectedcondition}, \ref{selectedcondviol}). Another (``democratic'') version, which treats all mixing operators (in the three-point functions) on an equal footing is given in Eqs. (\ref{selectcondition_demGIRS}, \ref{selectcondition_demGIRS_viol}). For all these variants of GIRS, we have extracted the next-to-leading-order conversion matrices between GIRS and $\overline{\rm MS}$. The conversion matrices, along with the lattice mixing matrices in GIRS, calculated nonperturbatively, allow for the determination of the renormalized physical matrix elements of the four-quark operators in the $\overline{\rm MS}$ scheme.

An important direction for future investigation is the assessment of the reliability of the GIRS variants in lattice simulations, particularly in comparison with more conventional renormalization schemes. In a forthcoming study, we intend to extend the GIRS framework to the renormalization of four-quark operators that mix with lower-dimensional operators. This case requires a more refined treatment due to the presence of power-divergent mixings, which pose additional conceptual and computational challenges.

\begin{acknowledgments}
 We would like to thank Vittorio Lubicz for useful communication. The project is implemented under the programme of social cohesion ``THALIA 2021-2027'', co-funded by the European Union through the Cyprus Research and Innovation Foundation (RIF). 
\end{acknowledgments}

\bibliography{refs}

\end{document}